\documentclass[graybox, envcountchap,usenatbib]{svmult}

\usepackage[numbers]{natbib} 
\usepackage{mathptmx}        
\usepackage{helvet}          
\usepackage{courier}         

\usepackage{makeidx}         
\usepackage{graphicx}        
\usepackage{multicol}        
\usepackage[bottom]{footmisc}

\makeindex             

\def\gtsima{$\; \buildrel > \over \sim \;$}
\def\ltsima{$\; \buildrel < \over \sim \;$}
\def\gtrsim{\lower.5ex\hbox{\gtsima}}
\def\lesssim{\lower.5ex\hbox{\ltsima}}

\begin{document}


\title{Formation channels of single and binary stellar-mass black holes}
\author{Michela Mapelli}
\institute{Michela Mapelli \at Dipartimento di Fisica e Astronomia Galileo Galilei, Vicolo dell'Osservatorio 3, I-35122, Padova, Italy \email{michela.mapelli@unipd.it}
\and \at INFN, Sezione di Padova, Via Marzolo 8, I-35122, Padova, Italy}
%
%
\maketitle

\abstract{These are exciting times for binary black hole (BBH) research. LIGO and Virgo detections are progressively drawing a spectacular fresco of BBH masses, spins and merger rates. In this review, we discuss the main formation channels of BBHs from stellar evolution and dynamics. Uncertainties on massive star evolution (e.g., stellar winds, rotation, overshooting and nuclear reaction rates), core-collapse supernovae and pair instability still hamper our comprehension of the mass spectrum and spin distribution of black holes (BHs), but substantial progress has been done in the field over the last few years. On top of this, the efficiency of mass transfer in a binary system and the physics of common envelope substantially affect the final BBH demography. Dynamical processes in dense stellar systems can trigger the formation of BHs in the mass gap and  intermediate-mass BHs via hierarchical BH mergers and via multiple stellar collisions. Finally, we discuss the importance of reconstructing the cosmic evolution of BBHs.
}


\section{Introduction: Observational facts about gravitational waves}

On 2015 September 14, the LIGO interferometers \cite{LIGOdetector} captured the gravitational wave (GW) signal from a binary black hole (BBH) merger \cite{abbottGW150914}. This event, named GW150914, is the first direct detection of GWs, about hundred years after Einstein's prediction. Over the last five years, LIGO and Virgo \cite{Virgodetector} witnessed a rapidly growing number of GW events: the second gravitational wave transient catalogue (GWTC-2, \cite{abbottO3a,abbottO3apopandrate,abbottO3aGR}) consists of 50 binary compact object mergers from the first (O1), the second (O2) and the first part of the third observing run (O3a) of the LIGO--Virgo collaboration (LVC).
Based on the results of independent pipelines, \cite{zackay2019}, \cite{udall2019}, \cite{venumadhav2020}  and \cite{nitz2020} claimed several additional GW candidates from O1 and O2. Furthermore, several dozens of public triggers from the second part of the third observing run (O3b) can be retrieved at \url{https://gracedb.ligo.org/}. 

Among the aforementioned detections, GW170817, the first binary neutron star (BNS) merger detected during O2, is the first and the only GW event unquestionably associated with an electromagnetic counterpart to date \cite{abbottmultimessenger,abbottGRB,goldstein2017,savchenko2017,margutti2017,coulter2017,soares-santos2017,chornock2017,cowperthwaite2017,nicholl2017,pian2017,alexander2017}.


Several other exceptional events were observed during O3a: the first unequal-mass BBH, GW190412 \cite{abbottGW190412}; ii) the second BNS, GW190425 \cite{abbottGW190425}; iii) the first black hole (BH) -- neutron star (NS) candidate, GW190814 \cite{abbottGW190814}, and iv) GW190521, which is the most massive system ever observed with GWs \cite{abbottGW190521,abbottGW190521astro}. Also, GW190521 is the first BBH event with a possible electromagnetic counterpart \cite{graham2020}. This growing sample represents a ``Rosetta stone'' to unravel the formation of binary compact objects.

\vspace{0.5cm}

Astrophysicists have learned several revolutionary concepts about compact objects from GW detections. Firstly, GW150914 has confirmed the existence of BBHs, i.e. binary systems composed of two BHs. BBHs have been predicted a long time ago (e.g. \cite{tutukov1973,thorne1987,schutz1989,kulkarni1993,sigurdsson1993,bethe1998,portegieszwart2000,colpi2003,belczynski2004}), but GW150914 is their first observational confirmation  \cite{abbottastrophysics}. Secondly, GW detections show that a number of BBHs are able to merge within a Hubble time. Thirdly, GW170817 has confirmed the connection between short gamma-ray bursts, kilonovae and BNS mergers \cite{abbottmultimessenger}.

Finally, most of the LIGO--Virgo BHs observed so far host BHs with mass in excess of 20 M$_\odot$. The very first detection, GW150914, has component masses equal to $m_1=35.6^{+4.7}_{-3.1}$ M$_\odot$ and $m_2=30.6^{+3.0}_{-4.4}$ M$_\odot$ \cite{abbottO1}. This was a genuine surprise for the astrophysicists \cite{abbottastrophysics}, because the only stellar BHs for which we have a dynamical mass measurement, i.e. about a dozen of BHs in X-ray binaries, have  mass  $\leq{}20$~M$_{\odot}$ \cite{orosz2003,ozel2010,miller2021}. Moreover, most theoretical models available five years ago did not predict the existence of BHs with mass $m_{\rm BH}>30$ M$_\odot$ (but see \cite{woosley2002,heger2002,mapelli2009,mapelli2010,belczynski2010,fryer2012,mapelli2013,ziosi2014,spera2015} for a few exceptions). 
Thus, the first GW detections have urged the astrophysical community to deeply revise the models of BH formation and evolution.

The recently published O3a events add complexity to this puzzle. 
In particular, the secondary component of GW190814, with a mass  $m_2=2.59^{+0.08}_{-0.09}$ M$_\odot$ \cite{abbottGW190814}, is either the lightest BH or the most massive NS ever observed, questioning the proposed existence of a mass gap between 2 and 5 M$_\odot$ \cite{ozel2010,farr2011}.

The merger product of GW190521 ($m_{\rm f}=142^{+28}_{-16}$ M$_\odot$, \cite{abbottGW190521,abbottGW190521astro}) is the first intermediate-mass BH (IMBH) ever detected by LIGO and Virgo and is the first BH with mass in the range $\sim{}100-1000$ M$_\odot$ ever observed not only with GWs but also in the electromagnetic spectrum. Moreover, the mass of the primary component, $m_1=85^{+21}_{-14}$ M$_\odot$, falls inside the predicted pair instability mass gap, as we will discuss later in this review. According to the re-analysis of \cite{nitzcapano2021} (see also \cite{fishbach2020}), GW190521 could be an intermediate-mass ratio inspiral with primary (secondary) mass $168^{+15}_{-61}$ M$_\odot$ ($16^{+33}_{-3}$ M$_\odot$), when assuming a uniform in mass-ratio prior. This interpretation avoids a violation of the mass gap, but requires the formation of an IMBH to explain the primary component. 

\vspace{0.5cm}

LIGO and Virgo do not observe only masses, they also allow us to extract information on spins and merger rates. The local BBH merger rate density inferred from GWTC-2 is $23.9_{-8.6}^{+14.9}$ ($58_{-29}^{+54}$) Gpc$^{-3}$ yr$^{-1}$ within the 90\% credible interval, when GW190814 is included in (excluded from) the sample of BBHs \cite{abbottO3apopandrate}. 
The LVC also estimated an upper limit for the merger rate density of BH--NS binaries ($\mathcal{R}_{\rm BHNS}<610$ Gpc$^{-3}$ yr$^{-1}$, \cite{abbottO2}) and inferred a BNS merger rate density $\mathcal{R}_{\rm BNS}=320^{+490}_{-240}$ Gpc$^{-3}$ yr$^{-1}$ within the 90\% credible interval \cite{abbottO3apopandrate}.

The uncertainties on the spins of each binary component are still too large to draw strong conclusions, apart from very few cases (for example, the spin of the primary component of GW190814 is close to zero, see Figure~6 of \cite{abbottGW190814}). However, LIGO and Virgo allow to give an estimate of two spin combinations, which are called effective spin ($\chi_{\rm eff}$) and precessing spin ($\chi_{\rm p}$). The effective spin is defined as
\begin{equation}\label{eq:chieff}
  \chi_{\rm eff}=\frac{(m_1\,{}\vec{\chi}_1+m_2\,{}\vec{\chi}_2)}{m_1+m_2}\cdot{}\frac{\vec{L}}{L},
\end{equation}
where $m_1$ and $m_2$ ($\vec{\chi}_1$ and $\vec{\chi}_2$) are the masses (dimensionless spin parameters) of the primary and secondary component of the binary, respectively and $\vec{L}$ is the Newtonian orbital angular momentum vector of the binary. The dimensionless spin parameters are defined as $\chi_i\equiv{}S_i\,{}c/(G\,{}m_i^2)$, where $S_i$ is the magnitude of the spin vector, $c$ the speed of light and $G$ the gravity constant. Hence, the effective spin can take any values between $-1$ and 1, where $\chi_{\rm eff}=1$ ($-1$) means that the two BHs are both maximally rotating and perfectly aligned (anti-aligned) with respect to the orbital angular momentum of the BBH, while $\chi_{\rm eff}=0$ indicates either that the two BHs are perfectly non-rotating or that both spins lie in the plane of the BBH orbit.

The precessing spin is defined as
\begin{equation}
\chi_{\rm p}=\frac{c}{B_1\,{}G\,{}m_1^2}\,{}\max{(B_1\,{}S_{1\perp{}},\,{}B_2\,{}S_{2\,{}\perp})},  
\end{equation}
where $B_1=2+3\,{}q/2$ and $B_2=2+3/(2\,{}q)$, $q=m_2/m_1\leq{}1$, $S_{1\perp}$ and $S_{2\perp}$ are the components of the spin vectors perpendicular to the orbital angular momentum. Hence, $\chi_{\rm p}$ can take values between 0 (no spin components in the orbital plane) and 1 (at least one spin being maximal and lying in the orbital plane). 
$\chi_{\rm p}$ is called precessing spin because spin components misaligned with respect to the orbital angular momentum of the binary drive precession \cite{schmidt2015}.

In current GW observations, small values of $\chi_{\rm eff}$ are preferred and $\chi_{\rm p}$ is unconstrained \cite{abbottO2,abbottO2popandrate}, with a few exceptions. GW151226, GW170729, GW190412 and GW190425, together with a few candidate events (e.g., GW190517\_055101) show support for positive values of $\chi_{\rm eff}$ \cite{abbottO2,abbottGW190412,abbottO3a}. If we look at the overall population of GWTC-2 BBHs \cite{abbottO3apopandrate}, $\sim{}12$\% to 44\% of BBH systems have spins tilted by more than 90$^\circ$ with respect to their orbital angular momentum, supporting a negative effective spin parameter.

The precessing spin of GW190814  has a strong upper bound $\chi_{\rm p}<0.07$ \cite{abbottGW190814}. Finally, GW190521 shows mild evidence for a non-zero precessing spin ($\chi_{\rm p}=0.68_{-0.37}^{+0.25}$ within the 90\% credible interval, \cite{abbottGW190521astro}). In contrast, spin measurements in $X-$ray binaries point to a range of spin magnitudes, including high spins \cite{millermiller2015,miller2021}.

This review discusses the formation channels of BHs and BBHs in light of the challenges posed by recent GW detections. This field has witnessed an exponential growth of publications and models in the last few years. While I will try to give an overview as complete as possible of the main models and connected issues, it would be impractical to mention every interesting study in the span of this review.

\section{The formation of compact remnants from single stellar evolution and supernova explosions}

Black holes (BHs) and neutron stars (NSs) are expected to form as remnants of massive ($\gtrsim{}8$ M$_\odot$) stars. An alternative theory predicts that BHs can also form  from gravitational collapse in the early Universe (the so called primordial BHs, e.g. \cite{carr1974,bird2016,carr2016,inomata2016}). In this review, we will focus on BHs of stellar origin.

The mass function of BHs is highly uncertain, because it may be affected by a number of barely understood processes. In particular, stellar winds and supernova (SN) explosions both play a major role on the formation of compact remnants. Processes occurring in close binary systems (e.g. mass transfer and common envelope) are a further complication and will be discussed in the next Section.  

\subsection{Stellar winds and stellar evolution}

Stellar winds are outflows of gas from the atmosphere of a star. In cold stars (e.g. red giants and asymptotic giant branch stars) they are mainly induced by radiation pressure on dust, which forms in the cold outer layers (e.g. \cite{vanloon2005}). In massive hot stars (O and B main sequence stars, luminous blue variables and Wolf-Rayet stars), stellar winds are powered by the coupling between the momentum of photons and that of metal ions present in the stellar photosphere. A large number of strong and weak resonant metal lines are responsible for this coupling (see e.g. \cite{kudritzki2000} for a review).

Understanding stellar winds is tremendously important for the study of compact objects, because mass loss determines the pre-SN mass of a star (both its total mass and its core mass), which in turn affects the outcome of an SN explosion \cite{fryer1999,fryer2001,mapelli2009,mapelli2010,belczynski2010}. 

Early work on stellar winds (e.g. \cite{abbott1982,kudritzki1987,leitherer1992}) highlighted that the mass loss of O and B stars depends on metallicity as $\dot{m}\propto{}Z^\alpha$ (with $\alpha{}\sim{}0.5-1.0$, depending on the model). However, such early work did not account for multiple scattering, i.e. for the possibility that a photon interacts several times before being absorbed or leaving the photosphere. Vink et al. (2001, \cite{vink2001}) accounted for multiple scatterings and found a  universal metallicity dependence $\dot{m}\propto{}Z^{0.85}\,{}v_\infty^p$, where $v_\infty$ is the terminal velocity and $p=-1.23$ ($p=-1.60$) for stars with effective temperature $T_{\rm eff}\gtrsim{}25000$ K ($12000\,{}{\rm K}\lesssim{}T_{\rm eff}\lesssim{}25000$~K).

The situation is more uncertain for post-main sequence stars. For Wolf-Rayet (WR) stars, i.e. naked helium cores, \cite{vinkdekoter2005} predict a similar trend with metallicity $\dot{m}\propto{}Z^{0.86}$. With a different numerical approach (which accounts also for wind clumping), \cite{graefener2008} find a strong dependence of WR mass loss on metallicity but also on the electron-scattering Eddington factor $\Gamma_e=\kappa_e\,{}L\,{}/(4\,{}\pi{}\,{}c\,{}G\,{}m)$, where $\kappa_e$ is the cross section for electron scattering, $L$ is the stellar luminosity, $c$ is the speed of light, $G$ is the gravity constant, and $m$ is the stellar mass. The importance of $\Gamma_e$ has become increasingly clear in the last few years \cite{graefener2011,vink2011,vink2016}, but only few stellar evolution models include this effect.

\begin{figure}
\center{
\includegraphics[width=12cm]{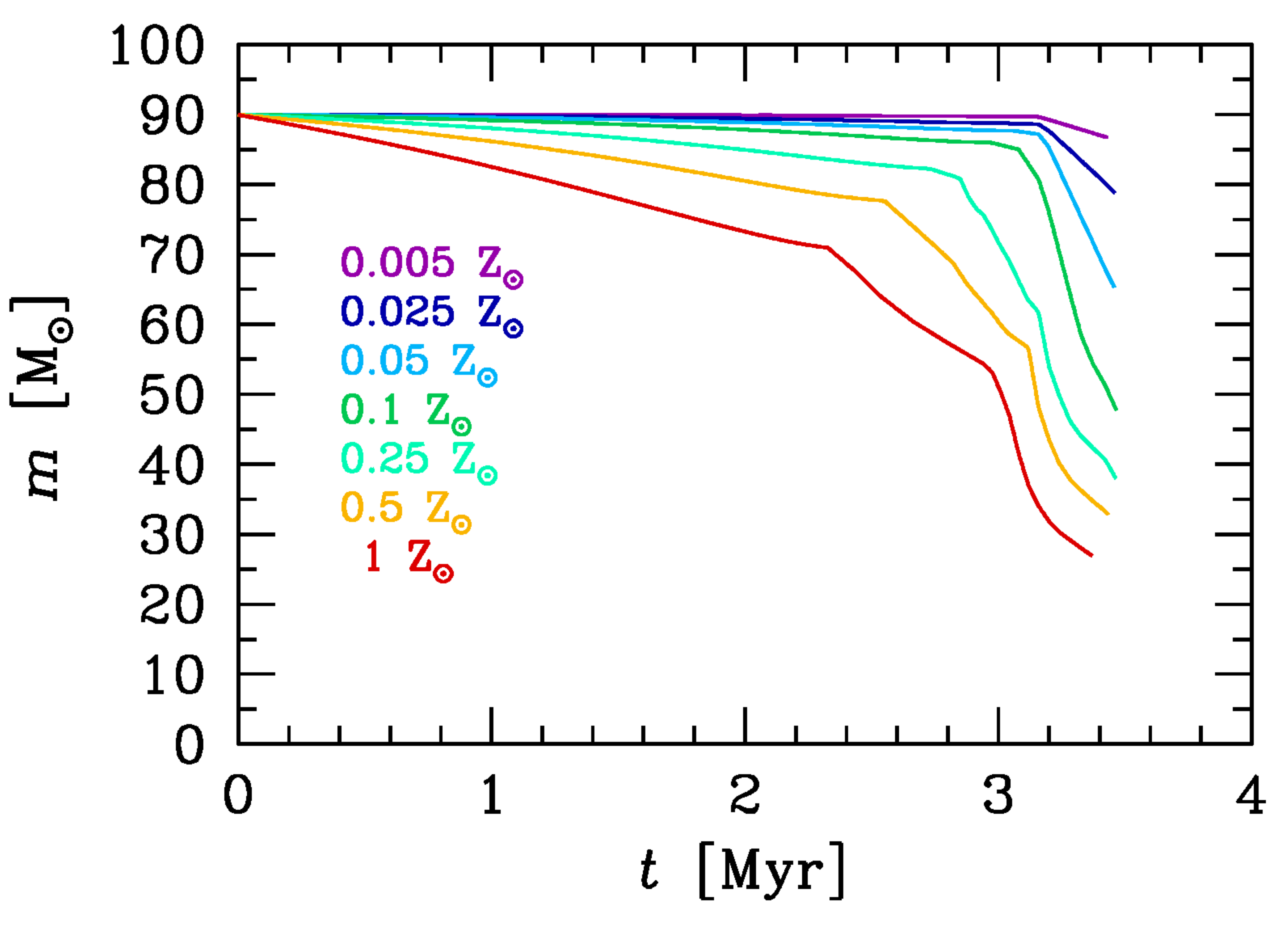}     
}
\caption{\label{fig:massloss}Evolution of stellar mass as a function of time for a star with ZAMS mass $m_{\rm ZAMS}=90$~M$_\odot$ and seven different metallicities, ranging from 0.005~Z$_\odot{}$ up to Z$_\odot$ (we assumed $Z_\odot=0.02$). These curves were obtained with the {\sc SEVN} population-synthesis code \cite{spera2015}, adopting {\sc PARSEC} stellar evolution tracks \cite{chen2015}.}
\end{figure}

For example, \cite{tang2014,chen2015} adopt a mass loss prescriptions for massive hot stars (O and B stars, luminous blue variables and WR stars) that scales as $\dot{m}\propto{}Z^\alpha$, where $\alpha{}=0.85$ if $\Gamma_e<2/3$, $\alpha{}=2.45-2.4\,{}\Gamma_e$ if $2/3\leq{}\Gamma_e\leq{}1$ and $\alpha=0.05$ if $\Gamma_e>1$. This simple formula accounts for the fact that metallicity dependence tends to vanish when the star is close to be radiation pressure dominated, as clearly shown by figure 10 of \cite{graefener2008}. Figure~\ref{fig:massloss} shows the mass evolution of a star with zero-age main sequence (ZAMS) mass $m_{\rm ZAMS}=90$ M$_\odot$ for seven different metallicities, as obtained with the {\sc SEVN} code \cite{spera2015}. At the end of its life, a solar-metallicity star (here we assume $Z_\odot=0.02$) has lost more than $2/3$ of its initial mass, while the most metal-poor star in the Figure ($Z=0.005$ Z$_\odot$) has retained almost all its initial mass.

Other aspects of massive star evolution also affect the pre-SN mass of a star. For example, surface magnetic fields appear to strongly quench stellar winds by magnetic confinement \cite{georgy2017,petit2017,keszthelyi2019}. 
In particular, \cite{petit2017} show that a non-magnetic star model with metallicity $\sim{}0.1$ Z$_\odot$ and a magnetic star model with solar metallicity and Alfv\'en radius $R_A\sim{}4\,{}R_\odot$ undergo approximately the same mass loss according to this model. This effect cannot be neglected because surface magnetic fields are detected in $\sim{}10$ per cent of the hot stars \cite{wade2016}, but is currently not included in models of compact-object formation.


Finally,  rotation affects the evolution of a massive star in several ways (e.g. \cite{maeder2009,chieffi2013,limongi2017,limongi2018,chieffi2020}). 
As a general rule of thumb, rotation increases the stellar luminosity. This implies that mass loss is generally enhanced if rotation is accounted for. On the other hand, rotation also induces chemical mixing, which leads to the formation of larger helium and carbon-oxygen cores. While enhanced mass loss implies smaller pre-SN masses, the formation of bigger cores has strong implications for the final fate of a massive star, as we discuss in the following Section.

\subsection{Core-collapse supernova (SN) or direct collapse}

Whether a star undergoes a successful core-collapse SN  or a failed SN is the first key question to address, in order to assess the properties of the final compact object. A star undergoing a successful core-collapse SN explosion will leave a NS or a light BH, while stars that end their life with a failed SN will become rather massive BHs ($>20$ M$_\odot$), because most (if not all) of the final mass of the star collapses into a BH directly. Addressing this question is a challenge for several reasons.

The mechanisms triggering iron core-collapse SNe are still highly uncertain. The basic framework and open issues are the following. As the mass of the central degenerate core reaches the Chandrasekhar mass \cite{chandrasekhar1931}, the degeneracy pressure of relativistic electrons becomes insufficient to support it against collapse. Moreover, electrons are increasingly removed, because protons capture them producing neutrons and neutrinos. This takes the core into a new state, where matter is essentially composed of neutrons, which support the core against collapse by their degeneracy pressure. To reach this new equilibrium, the core collapses from a radius of few thousand km down to a radius of few ten km in less than a second.  The gravitational energy gained from the collapse  is $W\sim{}5\times{}10^{53}\,{} {\rm erg} \,{}(m_{\rm PNS}/1.4\,{}M_\odot)^2\,{}(10\,{}{\rm km}/R_{\rm PNS})$, where $m_{\rm PNS}$ and $R_{\rm PNS}$ are the mass and radius of the proto-neutron star (PNS). 

The main problem is to explain how this gravitational energy can be --at least partially-- transferred to the stellar envelope triggering the SN explosion \cite{colgatewhite1966,bethewilson1985}. 
Several mechanisms have been proposed, including rotationally-driven SNe and/or magnetically-driven SNe (see e.g. \cite{janka2012,foglizzo2015} and references therein). The most commonly investigated mechanism is the convective SN engine (see e.g. \cite{fryer2012}). According to this model, the collapsing core drives a bounce shock. For the SN explosion to occur, this shock must reverse the supersonic infall of matter from the outer layers of the star. Most of the energy in the shock consists in a flux of neutrinos. As soon as neutrinos are free to leak out (because the shock has become diffuse enough), their energy is lost and the shock stalls. The SN occurs only if the shock is revived by some mechanism. In the convective SN scenario, the region between the PNS surface and the shock stalling radius can become convectively unstable (e.g. because of a Rayleigh–Taylor instability). Such convective instability can  convert  the  energy  leaking out of the PNS in the form of neutrinos to kinetic energy pushing the convective region outward. If the convective region overcomes the ram pressure of the infalling material, the shock is revived and an explosion is launched. If not, the SN fails.

While this is the general idea of the convective engine, fully self-consistent simulations of core collapse with a state-of-the-art treatment of neutrino transport do not lead to explosions in spherical symmetry except for the lighter SN progenitors ($\lesssim{}10$ M$_\odot$, \cite{foglizzo2015,ertl2016}). Simulations which do not require the assumption of spherical symmetry (i.e. run at least in 2D) appear to produce successful explosions from first principles for a larger range of progenitor masses (see e.g. \cite{mullerjanka2012a,mullerjanka2012b}). However, 2D and 3D simulations  are still computationally challenging and cannot be used to make a study of the mass distribution of compact remnants.

Thus, in order to study compact-object masses, SN explosions are artificially induced by injecting in the pre-SN model some amount of kinetic energy (kinetic bomb) or thermal energy (thermal bomb) at an arbitrary mass location. The evolution of the shock is then followed by means of 1D hydrodynamical simulations with some relatively simplified treatment for neutrinos. This allows to simulate hundreds of stellar models.

Following this approach, O'Connor \& Ott (2011, \cite{oconnor2011}) propose a criterion to decide whether a SN is successful or not, based on the compactness parameter:
\begin{equation}\label{eq:compac}
\xi{}_m=\frac{m/{\rm M}_\odot}{R(m)/1000\,{}{\rm km}},
\end{equation}              
where $R(m)$ is the radius which encloses a given mass $m$. Usually, the compactness is defined for $m=2.5$ M$_\odot$ ($\xi_{2.5}$). \cite{oconnor2011} measure the compactness at core bounce\footnote{\cite{ugliano2012} show that $\xi_{2.5}$ is not significantly different at core bounce or at the onset of collapse.} in their simulations and find that the larger $\xi_{2.5}$ is, the shorter is the time to form a BH (as shown in their Figure 6). This means that stars with a larger value of $\xi_{2.5}$ are more likely to collapse to a BH without SN explosion. 
The work by Ugliano et al. (2012, \cite{ugliano2012}) and Horiuchi et al. (2014, \cite{horiuchi2014}) indicate that the best threshold between exploding and non-exploding models is $\xi_{2.5}\sim{}0.2$.

Ertl et al. (2016, \cite{ertl2016}) indicate that a single criterion (e.g. the compactness) cannot capture the complex physics of core-collapse SN explosions. They introduce a two-parameter criterion based on 
\begin{equation}
M_4=\frac{m(s=4)}{{\rm M}_\odot}\quad{}{\rm and}\quad{}\mu_4=\left[\frac{dm/{\rm M}_\odot}{dR/1000\,{}{\rm km}}\right]_{s=4},
\end{equation}
where $M_4$ is the mass (at the onset of collapse) where the dimensionless entropy per baryon is $s=4$, and $\mu_4$ is the spatial derivative at the location of $M_4$. This choice is motivated by the fact that, in their 1D simulations, the explosion sets shortly after $M_4$ has fallen through the shock and well before the shell enclosing $M_4+0.3$ M$_\odot$ has collapsed. They show that exploding models can be distinguished from non-exploding models in the $\mu_4$ versus $M_4\,{}\mu_4$ plane (see their Figure 6) by a linear fit
\begin{equation}
y(x)=k_1\,{}x+k_2
\end{equation}
where $y(x)=\mu_4$, $x=M_4\,{}\mu_4$, and $k_1$ and $k_2$ are numerical coefficients which depend on the model (see Table 2 of \cite{ertl2016}). The reason of this behaviour is that $\mu{}_4$ scales with the rate of mass infall from the outer layers (thus the larger $\mu_4$ is, the lower the chance of the SN to occur), while $M_4\,{}\mu_4$ scales with the neutrino luminosity (thus the larger $M_4\,{}\mu_4$ is, the higher the chance of a SN explosion). Finally, \cite{ertl2016} find that fallback is quite inefficient ($<0.05$ M$_\odot$) when the SN occur.

The models proposed by O'Connor \& Ott (2011, \cite{oconnor2011}) and Ertl et al. (2016, \cite{ertl2016}, see also \cite{sukhbold2014,sukhbold2016,pejcha2015,ertl2020}) are sometimes referred to as the ``islands of explodability'' scenario, because they predict a non-monotonic behaviour of SN explosions with the stellar mass. This means, for example, that while a star with a mass $m=25$ M$_\odot$ and a star with a mass $m=29$ M$_\odot$ might end their life with a powerful SN explosion, another star with intermediate mass between these two (e.g. with a mass $m=27$ M$_\odot$) is expected to directly collapse to a BH without SN explosion. Thus, these models predict the existence of islands of explodability, i.e. ranges of mass where a star is expected to explode, surrounded by mass intervals in which the star will end its life with a direct collapse.

The models discussed so far depend on quantities ($\xi_{2.5}$, $M_4$, $\mu_4$) which can be evaluated no earlier than the onset of core collapse. Thus, stellar evolution models are required which integrate a massive star till the iron core has formed. This is prohibitive for most stellar evolution models (with few remarkable exceptions, e.g. FRANEC \cite{chieffi2013} and MESA \cite{paxton2015}).

Fryer et al. (2012, \cite{fryer2012}) propose a simplified approach (see also \cite{fryer1999,fryer2001,belczynski2008}). They suggest that the mass of the compact remnant depends mostly on two quantities: the carbon-oxygen core mass $m_{\rm CO}$ and the total final mass of the star $m_{\rm fin}$. In particular, $m_{\rm CO}$ determines whether the star will undergo a core-collapse SN or will collapse to a BH directly (namely, stars with $m_{\rm CO}>11$ M$_\odot$ collapse to a BH directly), whereas $m_{\rm fin}$ determines the amount of fallback on the PNS. In this formalism, the only free parameter is the time to launch the shock. The explosion energy is significantly reduced if the shock is launched $\gg{}250$ ms after the onset of the collapse ({\it delayed} SN explosion) with respect to an explosion launched in the first $\sim{}250$ ms ({\it rapid} SN explosion, \cite{fryer2012}).

While this approach is quite simplified with respect to other prescriptions, \cite{limongi2018} and \cite{chieffi2020} show that there is a strong correlation between the final carbon-oxygen mass and the compactness parameter $\xi{}_{2.5}$ at the onset of collapse, regardless of the rotation velocity of the progenitor star (see figure 1 of \cite{mapelli2020}). Thus, we can conclude that the simplified models by \cite{fryer2012} can effectively describe the overall trend of a collapsing star, although they do not take into account several details of the stellar structure at the onset of collapse.

Recently, \cite{pattonsukhbold2020} propose an interesting alternative approach. They integrate a large grid of naked carbon-oxygen (CO) cores to the onset of core collapse and estimate the explodability of each model with the compactness $\xi{}_{2.5}$ \cite{oconnor2011} and with the parameter $M_4$ \cite{ertl2020}. Naked CO cores are  faster and simpler to evolve than full stellar models (with hydrogen and helium) and are less sensitive to metallicity. They made available their grid of simulations for implementation into population-synthesis codes. Other works (e.g. \cite{clausen2015,mandelmuller2020a,mandelmuller2020b}) highlight the stochasticity of the final direct collapse or core-collapse SN. Finally, even the most advanced formalisms to derive the explodability of massive stars should be taken \emph{with a grain of salt}, because of the complexity of the processes involved in core collapse SNe and because of the simplifications still included in the models \cite{burrows2018}.

Even if we were in the conditions to tell if a given star undergoes a failed SN instead of a successful SN, this would not mean we can automatically infer the final mass of the compact object. In the case of a failed SN, the main uncertainty on the final compact object mass is represented by the fate of the envelope \cite{mapelli2020}. In fact, the envelope of a massive giant star is rather loosely bound and even a small energy injection can unbind a fraction of it.

Fernandez et al. (2018, \cite{fernandezquataert2018}) show that a $0.1-0.5$ M$_\odot$ neutrino emission during the PNS phase causes a decrease in the gravitational mass of the core, resulting in an outward going pressure wave (sound pulse) that steepens into a shock as it travels out through the star. This might cause the ejection of a fraction of the loosely bound stellar envelope. According to \cite{fernandezquataert2018}, the ejected mass is a monotonically decreasing function of the envelope compactness (Figure~6 of \cite{fernandezquataert2018}), defined as
\begin{equation}
  \xi_{\rm env}\equiv{}\frac{M_{\rm cc}/{\rm M}_\odot}{R_{\rm cc}/R_\odot},
\end{equation}
where $M_{\rm cc}$ and $R_{\rm cc}$ are the total mass and radius of the star at the onset of core-collapse. With this formalism, the ejected mass is up to a few M$_\odot$ for red super-giant stars, and 1 M$_\odot$ for more compact stars like blue super-giant stars and WR stars.



The possibility of a direct collapse is supported by observations. A survey conducted with the Large Binocular Telescope to find quietly disappearing stars \cite{kochanek2008,gerke2015}  reported evidence for the disappearance of a $\sim{}25$ M$_\odot$ red super-giant star \cite{adams2017}. In addition, surveys of SNe indicate a dearth of red super-giant progenitors with mass $>20$ M$_\odot$ associated with Type IIp SNe \cite{kochanek2014}.

\subsection{Pair instability and the mass gap}

If the helium core of a star grows above $\sim{}30$ M$_\odot$ and the core temperature is  $\gtrsim{}7\times{}10^8$ K at the end of carbon burning, the process of electron-positron pair production  becomes effective. It removes photon pressure from the core producing a sudden contraction of the carbon-oxygen core, before the formation of an iron core \cite{fowler1964,barkat1967,rakavy1967,woosley2017}.  For $m_{\rm He}>135$ M$_\odot$, the contraction cannot be reversed and the star collapses directly into a BH \cite{woosley2017}. If $135\gtrsim{}m_{\rm He}\gtrsim{}64$ M$_\odot$, the collapse triggers an explosive burning of heavier elements, especially oxygen and silicon. This leads to a pair instability SN (PISN): the star is completely disrupted, leaving no compact remnant \cite{heger2002}. For $64\gtrsim{}m_{\rm He}\gtrsim{}32$ M$_\odot$, pair production induces a series of pulsations of the core (pulsational pair instability), which trigger an enhanced mass loss \cite{woosley2017}. At the end of this instability phase, the star finds a new equilibrium and evolves towards core-collapse: a compact object with non-zero mass is produced, less massive than we would expect without pulsational mass loss. 

The main effect of (pulsational) pair instability is to open a gap in the mass spectrum of BHs between approximately $\sim{}50_{-10}^{+20}$ M$_\odot$ and $\sim{}120$ M$_\odot$ \cite{belczynski2016pair,woosley2017,woosley2019,spera2017,giacobbo2018a,giacobbo2018b,marchant2019,marchant2020,stevenson2019,vanson2020}. The large uncertainty on the edges of the mass gap is due to our poor understanding of the physics of massive stars. In particular, \cite{farmer2019} and \cite{farmer2020} integrate pure-He stars with {\sc mesa} \cite{paxton2011,paxton2013,paxton2015} and show that the uncertainties on the $^{12}$C$(\alpha{},\gamma{})^{16}$O reaction rate \cite{deboer2017} are responsible for a change in the lower-edge of the mass gap of more than $\sim{}15$ M$_\odot$.

Mapelli et al. (2020, \cite{mapelli2020}) show that assuming that the hydrogen envelope collapses to the final BH can lead to an increase of the lower-edge of the mass gap from $\sim{}40$  M$_\odot$ to $\sim{}65$ M$_\odot$. Finally, \cite{costa2020} combine the uncertainties on the  $^{12}$C$(\alpha{},\gamma{})^{16}$O reaction rate with the uncertainties on the collapse of the hydrogen envelope and find that the mass gap progressively reduces from $\sim{}80-150$ M$_\odot$ (for a  rate computed with the standard $^{12}$C$(\alpha{},\gamma{})^{16}$O reaction rate $-1\,{}\sigma{}$) to $\sim{}92-110$ M$_\odot$ (for a standard rate $-\,{}2\,{}\sigma{}$) and even disappears (for a standard rate $-3\,{}\sigma{}$) as an effect of convection and envelope undershooting. These uncertainties leave open the possibility that the primary mass of GW190521 is the result of stellar evolution \cite{costa2020,belczynski2020,tanikawa2020a,tanikawa2020b}.


\subsection{The mass of compact remnants}\label{sec:remnants}

The previous Sections suggest that our knowledge of the compact remnant mass is hampered by severe uncertainties, connected with both stellar winds and SNe. Thus, models of the mass spectrum of compact remnants must be taken \emph{with a grain of salt}. However, a few robust features can be drawn. 
\begin{figure}
\center{
\includegraphics[width=12cm]{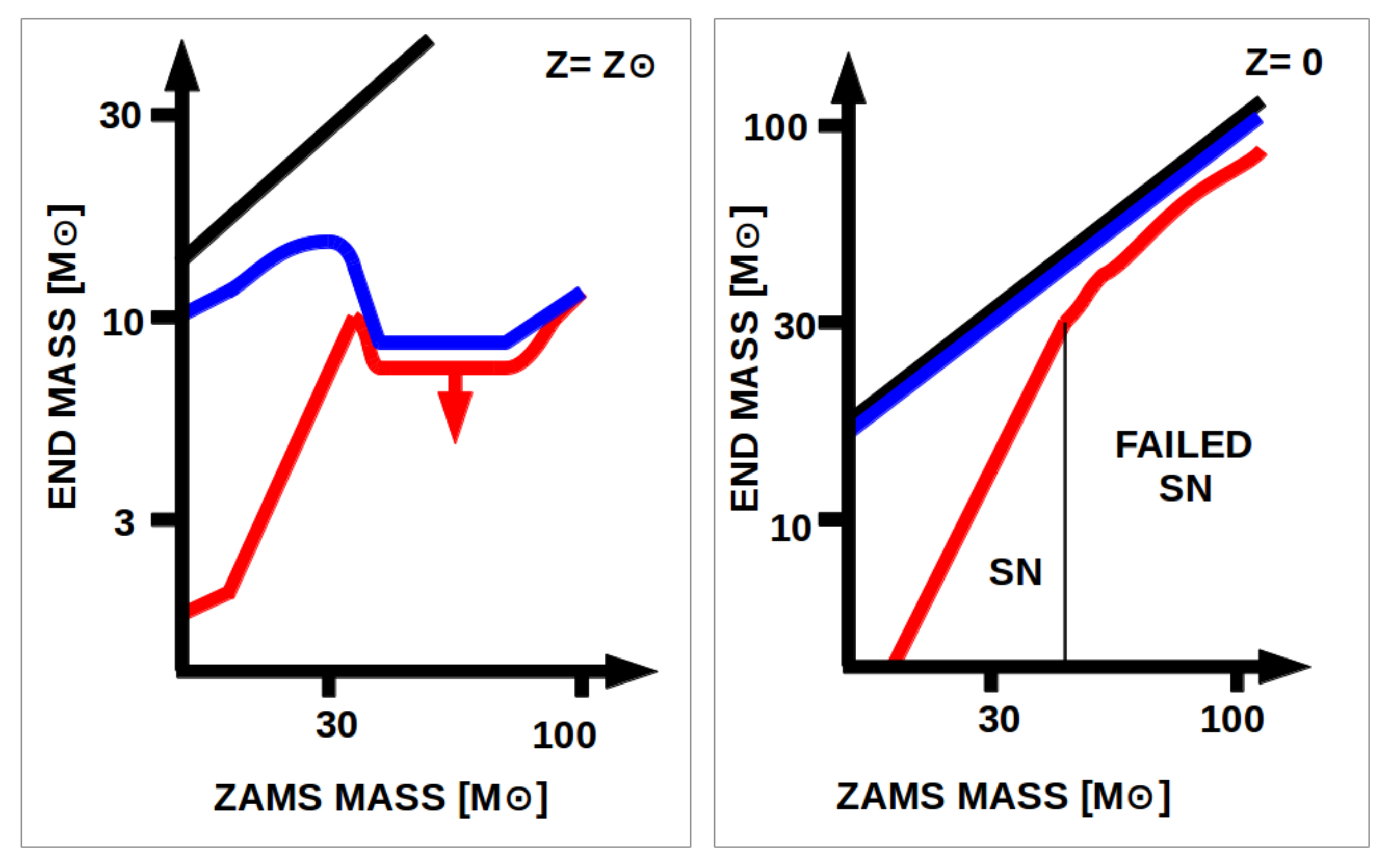}     
}
\caption{\label{fig:fromheger}Final mass of a star ($m_{\rm fin}$, blue lines) and mass of the compact remnant ($m_{\rm rem}$, red lines) as a function of the ZAMS mass of the star. The thick black line marks the region where $m_{\rm fin}=m_{\rm ZAMS}$. Left-hand panel: solar metallicity stars. Right-hand panel: metal-free stars.  The red arrow on the left-hand panel is an upper limit for the remnant mass. Vertical thin black line in the right-hand panel: approximate separation between successful and failed SNe at $Z=0$. This cartoon was inspired by Figures 2 and 3 of Heger et al. (2003 \cite{heger2003}).}
\end{figure}

Figure~\ref{fig:fromheger} is a simplified version of Figures 2 and 3 of Heger et al. (2003 \cite{heger2003}). The final mass of a star and the mass of the compact remnant are shown as a function of the ZAMS mass. The left and the right-hand panels show the case of solar metallicity and metal-free stars, respectively. In the case of solar metallicity stars, the final mass of the star is much lower than the initial one, because stellar winds are extremely efficient. The mass of the compact remnant is also much lower than the final mass of the star because a core-collapse SN always takes place. 

In contrast, a metal-free star (i.e., a population~III star) loses a negligible fraction of its mass by stellar winds (the blue and the black line in Figure~\ref{fig:fromheger} overlap). As for the mass of the compact remnant, Figure~\ref{fig:fromheger} shows that there are two regimes: below a given threshold ($\approx{}30-40$ M$_\odot$) the SN explosion succeeds even at zero metallicity and the mass of the compact remnant is relatively small. Above this threshold, the mass of the star (in terms of both core mass and envelope mass) is sufficiently large that the SN fails. Most of the final stellar mass collapses to a BH, whose mass is significantly larger than in the case of a SN explosion. The only exception is represented by the pair instability window: single metal-free stars with ZAMS mass $m_{\rm ZAMS}\sim{}140-260$ M$_\odot$ undergo a PISN and are completely destroyed, while single metal-free stars with  $m_{\rm ZAMS}\sim{}110-140$ M$_\odot$ undergo pulsational pair instability and leave smaller compact objects. In this simplified cartoon, we neglect the existence of islands of explodability.

What happens at intermediate metallicity between solar and zero? Predicting what happens to a metal-free star is relatively simple, because its evolution does not depend on the interplay between metals and stellar winds. The fate of a solar metallicity star is more problematic, because we must account for line-driven stellar winds, but most observational data about stellar winds are for nearly solar metallicity stars and allow us to calibrate our models for such high metallicity. 
In contrast, modelling intermediate metallicities is significantly more complicated, because the details depend on the interplay between metals and stellar winds and only limited data are available for calibration (mostly data for the Large and Small Magellanic Clouds, e.g. \cite{sander2019}).

As a rule of thumb (see e.g. \cite{fryer2012,spera2015}), we can draw the following considerations. If the zero-age main sequence (ZAMS) mass of a star is large ($m_{\rm ZAMS}\gtrsim{}30$ M$_\odot$), then the amount of mass lost by stellar winds is the main effect which determines the mass of the compact remnant. At low metallicity ($\lesssim{}0.1$ Z$_\odot$) and for a low Eddington factor ($\Gamma_e<0.6$), mass loss by stellar winds is not particularly large. Thus, the final mass $m_{\rm fin}$ and the carbon-oxygen mass $m_{\rm CO}$ of the star may be sufficiently large to avoid a core-collapse SN explosion: the star may form a massive BH ($\gtrsim{}20$ M$_\odot$) by direct collapse, unless a pair-instability or a pulsational-pair instability SN occurs. At high metallicity ($\approx{}$Z$_\odot$) or large Eddington factor ($\Gamma_e>0.6$), mass loss by stellar winds is particularly efficient and may lead to a small $m_{\rm fin}$ and $m_{\rm CO}$: the star is expected to undergo a core-collapse SN and to leave a relatively small remnant. 

If the ZAMS mass of a star is relatively low ($8<m_{\rm ZAMS}<30$ M$_\odot$), then stellar winds are not important (with the exception of asymptotic giant branch stars), regardless of the metallicity. In this case, the details of the SN explosion (e.g. energy of the explosion and amount of fallback) are crucial to determine the final mass of the remnant. This general sketch may be affected by several factors, such as pair-instability SNe, pulsational pair-instability SNe (e.g. \cite{woosley2017}) and an {\it island scenario} for core-collapse SNe (e.g. \cite{ertl2016}).

\begin{figure}
\center{
\includegraphics[width=12cm]{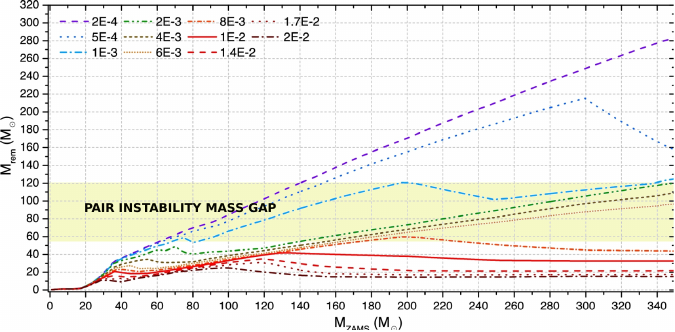}     
\includegraphics[width=12.1cm]{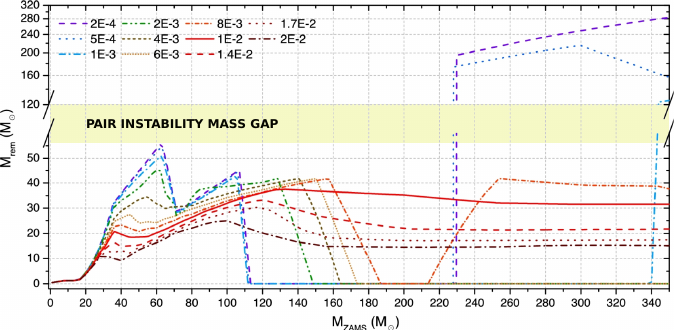}     
}
\caption{\label{fig:spera2017}Mass of the compact remnant ($m_{\rm rem}$) as a function of the ZAMS mass of the star ($m_{\rm ZAMS}$). Lower (upper) panel: pulsational pair-instability and pair-instability SNe are (are not) included. In both panels: dash-dotted brown line: $Z = 2.0\times{} 10^{-2}$; dotted dark orange line: $Z = 1.7\times{}10^{-2}$; dashed red line: $Z = 1.4\times{}10^{-2}$; solid red line: $Z = 1.0\times{} 10^{-2}$; short dash-dotted orange line: $Z = 8.0 \times{} 10^{-3}$; short dotted light orange line: $Z = 6.0 \times{} 10^{-3}$; short dashed green line: $Z = 4.0 \times{}10^{-3}$; dash-double dotted green line: $Z = 2.0 \times{} 10^{-3}$; dash-dotted light blue line: $Z = 1.0 \times{} 10^{-3}$; dotted blue line: $Z = 5.0\times{} 10^{-4}$; dashed violet line: $Z = 2.0\times{} 10^{-4}$. A delayed core-collapse SN mechanism has been assumed, following the prescriptions of \cite{fryer2012}. This Figure was adapted from Figures~1 and 2 of Spera \&{} Mapelli (2017, \cite{spera2017}).}
\end{figure}

The effect of pair-instability and pulsational pair-instability SNe is clearly shown in Figure~\ref{fig:spera2017}. The top panel was obtained accounting only for stellar evolution and core-collapse SNe. In contrast, the bottom panel also includes pair-instability and pulsational pair-instability SNe. This figure shows that the mass of the compact remnant strongly depends on the metallicity of the progenitor star if $m_{\rm ZAMS}\gtrsim{}30$ M$_\odot$. In most cases, the lower the metallicity of the progenitor is, the larger the maximum mass of the compact remnant \cite{heger2003,mapelli2009,belczynski2010,mapelli2010,mapelli2013,spera2015,spera2017}. However, for metal-poor stars  ($Z<10^{-3}$) with ZAMS mass $230>m_{\rm ZAMS}/{\rm M}_\odot>110$ pair instability SNe lead to the complete disruption of the star and no compact remnant is left. Only very massive ($m_{\rm ZAMS}>230$ M$_\odot$) metal-poor  ($Z<10^{-3}$) stars can collapse to a BH directly, producing intermediate-mass BHs (i.e, BHs with mass $\gtrsim{}100$ M$_\odot$).

If $Z<10^{-3}$ and  $110>m_{\rm ZAMS}\gtrsim{}60$ M$_\odot$, the star enters the pulsational pair-instability SN regime: mass loss is enhanced and the final BH mass is smaller ($m_{\rm BH}\sim{}30-55$ M$_\odot$, bottom panel of Fig.~\ref{fig:spera2017}) than we would have expected from direct collapse ($m_{\rm BH}\sim{}50-100$ M$_\odot$, top panel of Fig.~\ref{fig:spera2017}). 

Finally, the mass spectrum of relatively low-mass stars ($8<m_{\rm ZAMS}<30$ M$_\odot$) is not significantly affected by metallicity. The assumed core-collapse SN model is the most important factor in this mass range \cite{fryer2012}.

The models presented here do not take into account stellar rotation. Studied in \cite{limongi2018,mapelli2020,marchant2020}, stellar rotation produces larger cores because of chemical mixing. This shifts the minimum ZAMS mass for a star to undergo pulsational pair instability and PISN to lower values, because of the larger He core masses. The net result is a decrease of the maximum BH mass for fast-rotating stars ($\ge{}150$ km s$^{-1}$) with respect to low-rotating stars \cite{mapelli2020}.

\subsection{Compact object spins}

In the previous Sections, we have seen that the connection between the mass of a BH and the properties of its progenitor star is still highly uncertain. Our knowledge on the origin of BH spins is even more uncertain. 
It is reasonable to assume that a compact object inherits the spin of its progenitor (or at least of the core of its progenitor) if the progenitor collapses to a BH directly, without any SN explosion. In contrast, mass ejection during a SN explosion can significantly dissipate part of the final spin of the progenitor star. Hence, if the final spin of the progenitor star is not negligible, we would expect large birth spins for the most massive BHs, which form from direct collapse, and low birth spins for NSs and light BHs, which form from successful SN explosions.

Observations of Galactic pulsars seem to confirm the idea that NSs are born with relatively low $\chi{}$. Young pulsars (i.e., pulsars which did not have much time to slow down or to be recycled) have a
value of the spin parameter $\chi\lesssim{}0.01$ \cite{kramer2003,manchester2005}. This indicates that most of the spin of the progenitor was lost either before or during the SN. In contrast, millisecond pulsars are significantly spun up by mass accretion from a companion star: the fastest millisecond pulsar (B1937+21, \cite{hessels2006}) has $\chi{}\lesssim{}0.4$, but this value has almost nothing to do with the initial spin of the NS. 

On the other hand, the observed spins of BHs make us ``scratch our head''. Relatively low-mass BHs in high-mass X-ray binaries seem to show very large spins \cite{gou2009,gou2014,liu2008,reynolds2020,miller2021}. Considering that these BHs did not have enough time to spin up by mass accretion (for the short lifetime of their companion stars), this suggests large spins at birth. In contrast, LIGO--Virgo results support low values of $\chi_{\rm eff}$ for most of the BBHs \cite{abbottO2popandrate}, which can be interpreted either as strongly misaligned spins with respect to the binary angular momentum or as very low values of $\chi$.

From a theoretical perspective, the key question is then: what is the final spin of the progenitors of BHs and NSs? The final spin of a massive single star depends on mass loss and angular momentum transport. A mildly efficient transport by meridional currents (as adopted in, e.g., \cite{ekstrom2012,limongi2018}), leads to a non-negligible final spin of the star and to a high spin of the BH, if born from direct collapse \cite{belczynski2020,mapelli2020}. In contrast, \cite{fullerma2019} investigate efficient angular momentum transport via the magnetic Tayler-Spruit instability \cite{spruit2002,fuller2019} and find extremely slow spins ($\chi{}\sim{}10^{-2}$) for BHs born from single stars.

If the magnetic Tayler-Spruit instability is the dominant process for angular momentum transfer in massive stars, BH spins larger than $\chi{}\sim{}10^{-2}$ can be produced only by tidal torques in binary stars \cite{kushnir2016,qin2018,bavera2020,bavera2021} or by chemically homogeneous evolution \cite{maeder1987,woosley2006,yoon2006}. Overall, the initial rotation speed of very massive stars and the process of angular momentum transfer remain uncertain, hampering the predictive power of theoretical models on the spin distribution of BHs.

\subsection{Natal kicks}

Compact objects are expected to receive a natal kick from the parent SN explosion, because of asymmetries in the neutrino flux and/or in the ejecta (see \cite{janka2012} for a review). 
The natal kick has a crucial effect on the evolution of a binary compact object, because it can either unbind the binary or change its orbital properties. For example, a SN kick can increase the orbital eccentricity or misalign the spins of the two members of the binary.

Unfortunately, it is extremely difficult to quantify natal kicks from state-of-the-art SN simulations and measurements of natal kicks are scanty, especially for BHs. As to NSs, indirect observational estimates of SN kicks give contrasting results. Hobbs et al. (2005, \cite{hobbs2005}) found that a single Maxwellian with root mean square $\sigma{}_{\rm CCSN}=265$ km s$^{-1}$ can match the proper motions of 
73 young single pulsars in the Milky Way. Other works suggest a bimodal velocity distribution, with a first peak at low velocities (e.g. $\sim{}0$ km s$^{-1}$ according to \cite{fryer1998} or $\sim{}90$ km s$^{-1}$ according to \cite{arzoumanian2002}) and a second peak at high velocities ($>600$ km s$^{-1}$ according to \cite{fryer1998} or $\sim{}500$ km s$^{-1}$ for \cite{arzoumanian2002}). Similarly, Verbunt et al. (2017, \cite{verbunt2017}) indicate that a double Maxwellian distribution provides a significantly better fit to the observed velocity distribution than a single Maxwellian. Finally, the analysis of Beniamini \&{} Piran (2016, \cite{beniamini2016}) shows that low kick velocities ($\lesssim{}30$ km s$^{-1}$) are required to match the majority of Galactic BNSs, especially those with low eccentricity.

A possible interpretation of these observational results is that natal kicks depend on the SN mechanism (e.g. electron-capture versus core-collapse SN, e.g. \cite{giacobbo2019}) or on the binarity of the NS progenitor. For example, if the NS progenitor evolves in a close binary system (i.e., in a binary system where the two stars have exchanged mass with each other, see Section~\ref{sec:masstransfer}) it might undergo an ultra-stripped SN (see \cite{tauris2017} and references therein for more details). A star can undergo an ultra-stripped SN explosion only if it was heavily stripped by mass transfer to a companion \cite{tauris2013,tauris2015}.  
The natal kick of an ultra-stripped SN should be low \cite{tauris2015}, because of the small mass of the ejecta ($\lesssim{}0.1$ M$_\odot$). Low kicks ($\lesssim{}50$ km s$^{-1}$) for ultra-stripped core-collapse SNe are also confirmed by recent hydrodynamical simulations \cite{suwa2015,janka2017}.


As to BHs, the only indirect measurements of natal kicks arise from spatial distributions, proper motions and orbital properties of BHs in X-ray binaries (e.g. \cite{mirabel2017}). Evidence for a relatively small natal kick has been found for both GRO~J1655--40 \cite{willems2005} and Cygnus X-1 \cite{wong2012}, whereas H~1705--250 \cite{repetto2012,repetto2017} and XTE~J$1118+480$ \cite{mirabel2001,fragos2009} require high kicks  ($>100$ km s$^{-1}$). By analysing the position of BHs in X-ray binaries with respect to the Galactic plane, Repetto et al. (2012, \cite{repetto2012}) suggest that BH natal kicks should be as high as NS kicks.  Repetto et al. (2017, \cite{repetto2017}) perform a similar analysis but accounting also for binary evolution, and find that at least two BHs in X-ray binaries (H~1705--250 and XTE~J$1118+480$) require high kicks.

Most models of BBH evolution assume that natal kicks of BHs are drawn from the same distribution as NS kicks, but reduced by some factor. For example, linear momentum conservation suggests that
\begin{equation}
  v_{\rm BH}=\frac{\langle{}m_{\rm NS}\rangle{}}{m_{\rm BH}}\,{}v_{\rm NS},
\end{equation}
where $v_{\rm BH}$ is the natal kick of a BH with mass $m_{\rm BH}$, $\langle{}m_{\rm NS}\rangle{}\approx{1.33}\,{}{\rm M}_\odot$ is the average mass of NSs \cite{ozel2016} and $v_{\rm NS}$ is randomly drawn from the NS kick distribution (e.g., from a single or double Maxwellian). 

Alternatively, the natal kick can be reduced by the amount of fallback, under the reasonable assumption that fallback quenches the initial asymmetries. Following \cite{fryer2012},
\begin{equation}
  v_{\rm BH}=(1-f_{\rm fb})\,{}v_{\rm NS},
\end{equation}
where $f_{\rm fb}$ quantifies the fallback ($f_{\rm fb}=0$ for no fallback and $f_{\rm fb}=1$ for direct collapse). Most studies assume that  BHs born from direct collapse receive no kick \cite{fryer2012}. 

The model by \cite{giacobbo2020} can unify BH kicks and NS kicks, naturally accounting for ultra-stripped SNe and electron-capture SNe, which mostly lead to reduced kicks (e.g., \cite{suwa2015,tauris2017}). According to this toy model, the kick of a compact object can be described as
\begin{equation}
  v_k=f_{\rm H05}\,{}\frac{\langle{}m_{\rm NS}\rangle{}}{m_{\rm rem}}\,{}\frac{m_{\rm ej}}{\langle{}m_{\rm ej}\rangle{}},
\end{equation}
where  $m_{\rm rem}$ is the mass of the considered compact remnant, $m_{\rm ej}$ is the mass of the ejecta, $\langle{}m_{\rm ej}\rangle{}$ is the ejecta mass of a SN that leaves a single NS with mass $\langle{}m_{\rm NS}\rangle{}$, and $f_{\rm H05}$ is a number randomly drawn from a Maxwellian probability density curve with one-dimensional root-mean square velocity dispersion $\sigma{}_{\rm 1D}=265$ km s$^{-1}$ \cite{hobbs2005}. To derive this model, \cite{giacobbo2020} assume that the Maxwellian distribution fitted by \cite{hobbs2005} is a good proxy to the distribution of the kicks of NSs born from single stars and that kicks are the effect of asymmetries in mass ejecta ($\propto{}m_{\rm ej}$, see also \cite{bray2016,bray2018,tang2020}), modulated by linear momentum conservation ($\propto{}m_{\rm rem}^{-1}$). This simple formalism seems to solve the tension between the local merger rate of BNSs derived from gravitational-wave interferometers \cite{abbottO2popandrate,abbottGW190425} and the proper motions of Galactic young pulsars \cite{hobbs2005}.

In summary, natal kicks are one of the most debated issues about compact objects. Their actual amount has dramatic implications on the merger rate and on the properties (spin and mass distribution) of merging compact objects.

\section{Binaries of stellar black holes}

Naively, one could think that if two massive stars are members of a binary system, they will eventually become a BBH and the mass of each BH will be the same as if its progenitor star was a single star. This is true only if the binary system is sufficiently wide (detached binary) for its entire evolution. If the binary is tight enough, it will evolve through several processes which might significantly change its final fate. 

The so-called binary population-synthesis codes have been used to investigate the effect of binary evolution processes on the formation of BBHs in isolated binaries (e.g. \cite{portegieszwart1996,hurley2002,podsiadlowski2003,belczynski2008,mapelli2013,mennekens2014,eldridge2016,eldridge2017,mapelli2017,stevenson2017,giacobbo2018a,barrett2018,giacobbo2018b,giacobbo2019,kruckow2018,eldridge2019,spera2019,giacobbo2020,tanikawa2020a}). These are semi-analytic codes which combine a description of stellar evolution with prescriptions for SN explosions and with a formalism for binary evolution processes. In the following, we mention some of the most important binary-evolution processes and we briefly discuss their treatment in the most used population-synthesis codes.

\subsection{Mass transfer}\label{sec:masstransfer}

If two stars exchange matter to each other, it means they undergo a mass transfer episode. This might be driven either by stellar winds or by an episode of Roche-lobe filling. When a massive star loses mass by stellar winds, its companion might be able to capture some of this mass. This will depend on the amount of mass which is lost and on the relative velocity of the wind with respect to the companion star. Based on the Bondi \& Hoyle (1944, \cite{bondi1944}) formalism, Hurley et al. (2002, \cite{hurley2002}) describe the mean mass accretion rate by stellar winds as
\begin{equation}
\dot{m}_2=\frac{1}{\sqrt{1-e^2}}\,{}\left(\frac{G\,{}m_2}{v_{\rm w}^2}\right)^2\,{}\frac{\alpha_{\rm w}}{2\,{}a^2}\,{}\frac{1}{[1+(v_{\rm orb}/v_{\rm w})^2]^{3/2}}\,{}\left|\dot{m}_{\rm 1}\right|,
\end{equation}
where $e$ is the binary eccentricity, $G$ is the gravitational constant, $m_2$ is the mass of the accreting star, $v_{\rm w}$ is the velocity of the wind, $\alpha_{\rm w}\sim{}3/2$ is an efficiency constant, $a$ is the semi-major axis of the binary, $v_{\rm orb}=\sqrt{G\,{}(m_1+m_2)/a}$ is the orbital velocity of the binary ($m_1$ being the mass of the donor), and $\dot{m}_1$ is the mass loss rate by the donor. Since $|\dot{m}_1|$ is usually quite low ($|\dot{m}_1|<10^{-3}$ M$_\odot$ yr$^{-1}$) and $v_{\rm w}$ is usually quite high  ($>1000$ km s$^{-1}$ for a line-driven wind) with respect to the orbital velocity, this kind of mass transfer is usually rather inefficient. However, 
most of the observed high-mass X-ray binaries \cite{vandenheuvel1976,vandenheuvel2019} and in particular all the systems with a WR star companion \cite{esposito2015} are wind-fed systems.

Mass transfer by Roche lobe overflow is usually more efficient than wind accretion. The Roche lobe of a star in a binary system is a teardrop-shaped equipotential surface surrounding the star. The Roche lobes of the two members of the binary are connected in just one point, which is the Lagrangian L1 point. A widely used approximate formula for the Roche lobe is \cite{eggleton1983}
\begin{equation}\label{eq:rlobe}
R_{\rm L,1}=a\,{}\frac{0.49\,{}q^{2/3}}{0.6\,{}q^{2/3}+\ln{\left(1+q^{1/3}\right)}},
\end{equation}
where $a$ is the semi-major axis of the binary and $q=m_1/m_2$ ($m_1$ and $m_2$ are the masses of the two stars in the binary). This formula describes the Roche lobe of a star with mass $m_1$, while the corresponding Roche lobe of a star with mass $m_2$ ($R_{\rm L,2}$) is obtained by swapping the subscripts. A star overfills (underfills) its Roche lobe when its radius is larger (smaller) than the Roche lobe. If a star overfills its Roche lobe, a part of its mass flows toward the companion star which can accrete (a part of) it.


Mass transfer obviously changes the mass of the two stars in a binary, and thus the final mass of the compact remnants of such stars, but also the orbital properties of the binary. If mass transfer is non conservative, which is the most realistic case, it leads to an angular momentum loss, which in turn affects the semi-major axis. Recently, \cite{bouffanais2020} show that the hypothesis of a highly non-conservative mass transfer (mass accretion efficiency $f_{\rm MT}\leq{}0.5$) is in tension with LVC data, if we assume that all BBHs observed by the LVC form via isolated binary evolution.

A crucial information about Roche lobe overflow is whether it is stable or unstable and on which timescale. The most commonly used approach can be described as follows \cite{webbink1985,portegieszwart1996,tout1997,hurley2002,eggleton2006}. Let us  assume that the stellar radius and mass are connected by a simple relation $R\propto{}m^\zeta{}$. Thus, the variation of the donor's radius during Roche lobe is
\begin{equation}
  \frac{{\rm d}R_1}{{\rm d}t}=\frac{\partial{}R_1}{\partial{}t}+\zeta{}\,{}\frac{R_1}{m_1}\,{}\frac{{\rm d}m_1}{{\rm d}t}.
\end{equation}
In the above equation, the term $\frac{\partial{}R_1}{\partial{}t}$ is due to nuclear burning, while the term with $\zeta{}$ measures the adiabatic or thermal response of the donor star to mass loss. Note that $\frac{{\rm d}m_1}{{\rm d}t}$ is the mass loss from the donor; hence it is always negative.

Similarly, the change of the size of the Roche lobe of the donor $R_{\rm L,1}$ can be estimated as
\begin{equation}
  \frac{{\rm d}R_{\rm L,1}}{{\rm d}t}=\frac{\partial{}R_{\rm L,1}}{\partial{}t}+\zeta{}_{\rm L}\,{}\frac{R_{\rm L,1}}{m_1}\,{}\frac{{\rm d}m_1}{{\rm d}t},
\end{equation}
where $\frac{\partial{}R_{\rm L,1}}{\partial{}t}$ depends on tides and GW radiation, while $\zeta_{\rm L}$ describes the response of the Roche lobe to mass loss: the Roche lobe might shrink or expand. 
If $\zeta_{\rm L}>\zeta{}$, the Roche lobe shrinks faster than the radius of the star does and the mass transfer is unstable, otherwise it remains stable until the radius changes significantly by nuclear burning.

Mass transfer can be unstable either on a dynamical timescale (if $\zeta{}$ describes the adiabatic response of the donor and $\zeta{}<\zeta{}_{\rm L}$) or on a thermal timescale (if $\zeta{}$ describes the thermal response of the donor and $\zeta{}<\zeta{}_{\rm L}$). If mass transfer is dynamically unstable  or both stars overfill their Roche lobe, then the binary is expected to merge -- if the donor lacks a steep density gradient between the core and the envelope --, or to enter common envelope (CE) -- if the donor has a clear distinction between core and envelope.

\subsection{Common envelope (CE)}

If two stars enter CE, their envelope(s) stop co-rotating with their cores. The two stellar cores (or the compact object and the core of the companion star, if the binary is already single degenerate) are embedded in the same non-corotating envelope and start spiralling in as an effect of gas drag exerted by the envelope. Part of the orbital energy lost by the cores as an effect of this drag is likely converted into heating of the envelope, making it more loosely bound. If this process leads to the ejection of the envelope, then the binary survives, but the post-CE binary is composed of two naked stellar cores (or of a compact object and a naked stellar core). Moreover, the orbital separation of the two cores (or the orbital separation of the compact object and the stellar core) is considerably smaller than the initial orbital separation before the CE, as an effect of the spiral in\footnote{A short-period (from a few hours to a few days) binary system composed of a naked helium core and BH might be observed as an X-ray binary, typically a WR X-ray binary. In the local Universe, we know a few ($\sim{}7$) WR X-ray binaries, in which a compact object (BH or NS) accretes mass through the wind of the naked stellar companion (see e.g. \cite{esposito2015} for more details). These rare X-ray binaries are thought to be good progenitors of merging compact-object binaries.}. This circumstance is crucial for the fate of a BBH. In fact, if the binary which survives a CE phase evolves into a BBH, this BBH will have a short semi-major axis ($a\lesssim{}100$~R$_\odot$), much shorter than the sum of the maximum radii of the progenitor stars, and may be able to merge by GW emission within a Hubble time.

\begin{figure}
\center{
\includegraphics[width=12cm]{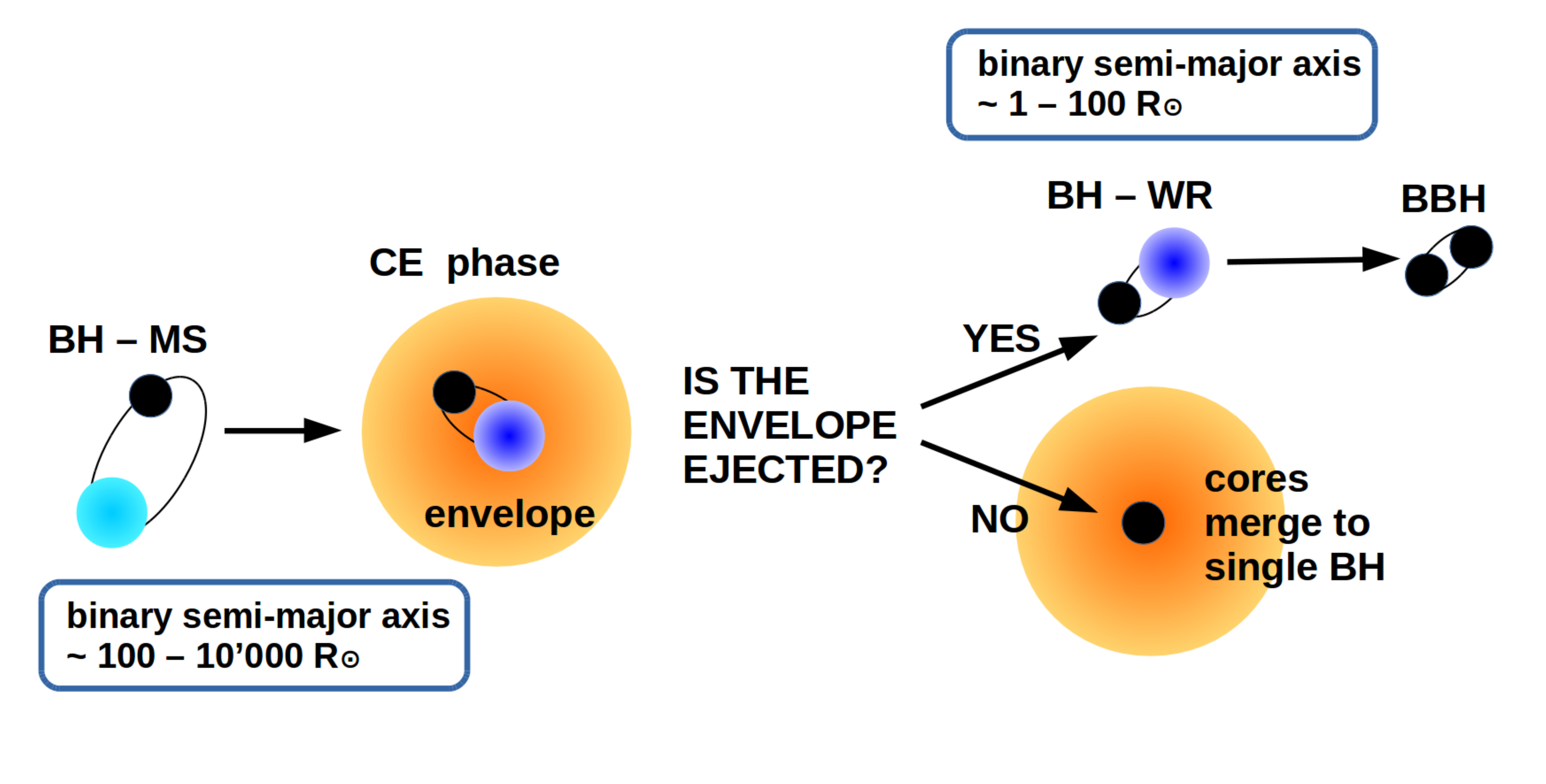} 
}
\caption{\label{fig:commonenv}Schematic representation of the evolution of a BBH  through CE. The companion of the BH is initially in the main sequence (MS). In the cartoon, the BH is indicated by a black circle, while the MS companion is indicated by the light blue circle. When the companion evolves off the MS, becoming a giant star, it overfills its Roche lobe. The BH and the giant star enter a CE (the CE is indicated in orange, while the core of the giant is represented by the dark blue circle). The core of the giant and the BH spiral in because of the gas drag exerted by the envelope. If the envelope is ejected, we are left with a new binary, composed of the BH and the naked helium core of the giant. The new binary has a much smaller orbital separation than the initial binary. If the naked helium core becomes a BH and its natal kick does not disrupt the binary, then a BBH is born, possibly with a small semi-major axis. In contrast, if the envelope is not ejected, the BH and the helium core spiral in, till they merge together. A single BH is left in this case.}
\end{figure}

In contrast, if the envelope is not ejected, the two cores (or the compact object and the core) spiral in till they eventually merge. This premature merger of a binary during a CE phase prevents the binary from evolving into a BBH. The cartoon in Figure~\ref{fig:commonenv} summarizes these possible outcomes.

The $\alpha{}$ formalism \cite{webbink1984} is the most common formalism adopted to describe a common envelope. The basic idea of this formalism is that the energy needed to unbind the envelope comes uniquely from the loss of orbital energy of the two cores during the spiral in. The fraction of the orbital energy of the two cores which goes into unbinding the envelope can be expressed as
\begin{equation}
\Delta{}E=\alpha{}\,{}(E_{\rm b,f}-E_{\rm b,i})=\alpha{}\,{}\frac{G\,{}m_{\rm c1}\,{}m_{\rm c2}}{2}\,{}\left(\frac{1}{a_{\rm f}}-\frac{1}{a_{\rm i}}\right),
\end{equation}
where $E_{\rm b,i}$ ($E_{\rm b,f}$) is the orbital binding energy of the two cores before (after) the CE phase, $a_{\rm i}$ ($a_{\rm f}$) is the semi-major axis before (after) the CE phase, $m_{\rm c1}$ and $m_{\rm c2}$ are the masses of the two cores, and $\alpha{}$ is a dimensionless parameter that measures which fraction of the removed orbital energy is transferred to the envelope. If the primary is already a compact object (as in Figure~\ref{fig:commonenv}), $m_{\rm c2}$ is the mass of the compact object.

The binding energy of the envelope is
\begin{equation}
E_{\rm env}=\frac{G}{\lambda}\,{}\left[\frac{m_{\rm env,1}\,{}m_{\rm 1}}{R_1}+\frac{m_{\rm env,2}\,{}m_{\rm 2}}{R_2}\right],
\end{equation}
where $m_1$ and $m_2$ are the masses of the primary and the secondary member of the binary, $m_{\rm env,1}$ and $m_{\rm env,2}$ are the masses of the envelope of the primary and the secondary member of the binary, $R_1$ and $R_2$ are the radii of the primary and the secondary member of the binary, and $\lambda{}$ is the parameter (or the function) which measures the concentration of the envelope (the smaller $\lambda{}$ is, the more concentrated is the envelope).

By imposing $\Delta{}E=E_{\rm env}$ we can derive the value of the final semi-major axis $a_{\rm f}$ for which the envelope is ejected. 
This means that the larger (smaller) $\alpha{}$ is, the larger (smaller) the final orbital separation. If the resulting $a_{\rm f}$ is lower than the sum of the radii of the two cores (or than the sum of the Roche lobe radii of the cores), then the binary will merge during CE, otherwise the binary survives. 

Actually, we have known for a long time (see \cite{ivanova2013} for a review) that this simple formalism is a poor description of the physics of CE, which is considerably more complicated. A healthy treatment of CE should take into account not only the orbital energy of the cores and the binding energy of the envelope, but also i) the thermal energy of the envelope, which is the sum of radiation energy and kinetic energy of gas particles \cite{han1994}, ii) the recombination energy (as the envelope expands it cools down, the plasma recombines and some atoms even form molecules, releasing binding energy, \cite{kruckow2016}), iii) tidal heating/cooling from stellar spin down/up \cite{ivanova2013}, iv) nuclear fusion energy \cite{ivanova2002}, v) the enthalpy of the envelope \cite{ivanova2011}, and vi) the accretion energy, which might drive outflows and jets \cite{soker2004NewA....9..399S,macleod2015ApJ...798L..19M,macleod2015ApJ...803...41M,soker2016NewAR..75....1S,macleod2017ApJ...838...56M,desoumi2020ApJ...897..130D}.

Moreover, the envelope concentration parameter $\lambda{}$ cannot be the same for all stars. It is expected to vary wildly not only from star to star but also during different evolutionary stages of the same star. Several authors \cite{xu2010,loveridge2011} have estimated $E_{\rm env}$ directly from their stellar models, 
significantly improving this formalism. However, even in this case, we cannot get rid of the $\alpha{}$ parameter. 


Thus, it is essential to model the physics of CE with analytic models and numerical simulations. A lot of effort has been put on this in the last few years, but the problem remains largely unconquered. Several recent studies investigate the onset of CE, when an unstable mass transfer prevents the envelope from co-rotating with the core and leads to the plunge-in of the companion inside the envelope \cite{macleod2017ApJ...835..282M,macleod2018ApJ...863....5M,macleod2020ApJ...895...29M,macleod2020ApJ...893..106M, vick2020arXiv200805476V}.

Several hydrodynamical simulations model the fast spiral in phase after plunge-in \cite{rickertaam2008,rickertaam2012,passy2012,ohlmann2016}, when the two cores spiral in on a dynamical time scale ($\approx{}100$ days). At the end of this dynamical spiral in only a small fraction of the envelope ($\sim{}25$\%, \cite{ohlmann2016}) appears to be ejected in most simulations. When the two cores are sufficiently close that they are separated only by a small gas mass, the spiral in slows down and the system evolves on the Kelvin-Helmholtz timescale of the envelope ($\approx{}10^{3-5}$ years).

Simulating the system for a Kelvin-Helmholtz timescale is prohibitive for current three-dimensional simulations (e.g., \cite{lawsmith2020}). Fragos et al. (2019, \cite{fragos2019}) reduce the complexity of the problem by simulating the entire CE evolution in just one dimension, with the hydrodynamic stellar evolution code  {\sc mesa} \cite{paxton2011,paxton2013,paxton2015}. They evolve a binary system composed of a NS and a 12 M$_\odot$ red super-giant star for a thermal timescale. In their model, envelope ejection is mostly driven by the thermal energy of the envelope and by the orbital energy, while recombination contributes to $\lesssim{}10$\% of the total energy required to eject the envelope. The the final system is a NS -- naked helium star system with an orbital separation of a few R$_\odot$, i.e. a good progenitor for a merging BNS. Their simulated system can be reproduced by the $\alpha{}$ formalism if $\alpha\approx{}5$. The apparently unphysical value of $\alpha{}>1$ is motivated by the fact that orbital energy is only one of the energy sources that participate in envelope ejection.

\subsection{Alternative evolution to CE}

Massive fast rotating stars can have a chemically homogeneous evolution (CHE): they do not develop a chemical composition gradient because of the mixing induced by rotation. This is particularly true if the star is metal poor, because stellar winds are not efficient in removing angular momentum. If a binary is very tight, the spins of its members are even increased during stellar life, because of tidal synchronisation. The radii of stars following CHE are usually much smaller than the radii of stars developing a chemical composition gradient \cite{demink2016,mandel2016}. This implies that even very tight binaries (few tens of solar radii) can avoid CE.

Marchant et al. (2016, \cite{marchant2016}) simulate tight binaries whose components are fast rotating massive stars. A number of their simulated binaries evolve into contact binaries where both binary components fill and even overfill their Roche volumes. If metallicity is sufficiently low and rotation sufficiently fast, these binaries may evolve as ``over-contact'' binaries: the over-contact phase differs from a classical CE phase because co-rotation can, in principle, be maintained as long as material does not overflow the L2 point. This means that 
spiral-in 
can be avoided, resulting in a stable system evolving on a nuclear timescale.

Such over-contact binaries maintain relatively small stellar radii during their evolution (few ten solar radii) and may evolve into a BBH with a very short orbital period. This scenario predicts the formation of merging BHs with relatively large masses ($>20$ M$_\odot$), nearly equal mass ($q=1$), and with large aligned spins (i.e., large $\chi_{\rm eff}$). While a large positive $\chi_{\rm eff}$ is not consistent with most of the LIGO--Virgo BBHs, the CHE model is a viable alternative to explain systems with large and aligned spins. Moreover, mass loss during the naked helium core phase might reduce the progenitor's spin and lead to lower BH spins than predicted by early models \cite{riley2021}. 

\subsection{BBH spins in the isolated binary evolution model}

Most evolutionary processes in an isolated binary star (tides, mass transfer, CE) lead to the alignment of the spins of the components to the orbital angular momentum of the binary \cite{hurley2002}. The only\footnote{Stegmann \&{} Antonini (2021, \cite{stegmann2021}) recently proposed a possible spin flip mechanism during mass transfer.} evolutionary process that can significantly misalign BH spins with respect to the orbital angular momentum is the SN explosion \cite{hurley2002,gerosa2013,rodriguez2016,gerosa2017,belczynski2020}.

Under the assumption that the system has no time to re-align spins between the first and the second SN and that SNe do not change the direction of compact-object spins but only the direction of the orbital angular momentum of the binary, we can derive the angle between the direction of the spins of the two compact objects  and that of the orbital angular momentum of the binary system as \cite{gerosa2013,rodriguez2016}
\begin{equation}
  \cos{\theta{}}=\cos{(\nu{}_1)}\,{}\cos{(\nu{}_2)}+\sin{(\nu{}_1)}\,{}\sin{(\nu{}_2)}\,{}\cos{(\phi{})},
\end{equation}
where $\nu{}_i$ (with $i=1,\,{}2$) is the angle between the new (${\bf L}_{\rm new}$) and the old (${\bf L}$) orbital angular momentum after a SN ($i=1$ corresponding to the first SN, $i=2$ corresponding to the second SN), so that $\cos{(\nu{})}=\hat{L}_{\rm new}\cdot{}\hat{L}_{\rm old}$, while $\phi$ is the phase of the projection of $\hat{L}$ into the orbital plane. As shown by \cite{rodriguez2016}, the most commonly adopted SN kick models fail producing a significant misalignment.

Hence, we expect that the isolated binary evolution model has a preference for BH spins aligned with the orbital angular momentum of the binary, especially in the case of massive BHs which undergo just a failed SN.

\subsection{Summary of the isolated binary formation channel}

In this Section, we have highlighted the most important aspects and the open issues of the {\emph{isolated binary formation scenario}}, i.e.  the model which predicts the formation of merging BHs through the evolution of isolated binaries. For isolated binaries we mean stellar binary systems which are not perturbed by other stars or compact objects.

\begin{figure}
\center{
\includegraphics[width=12cm]{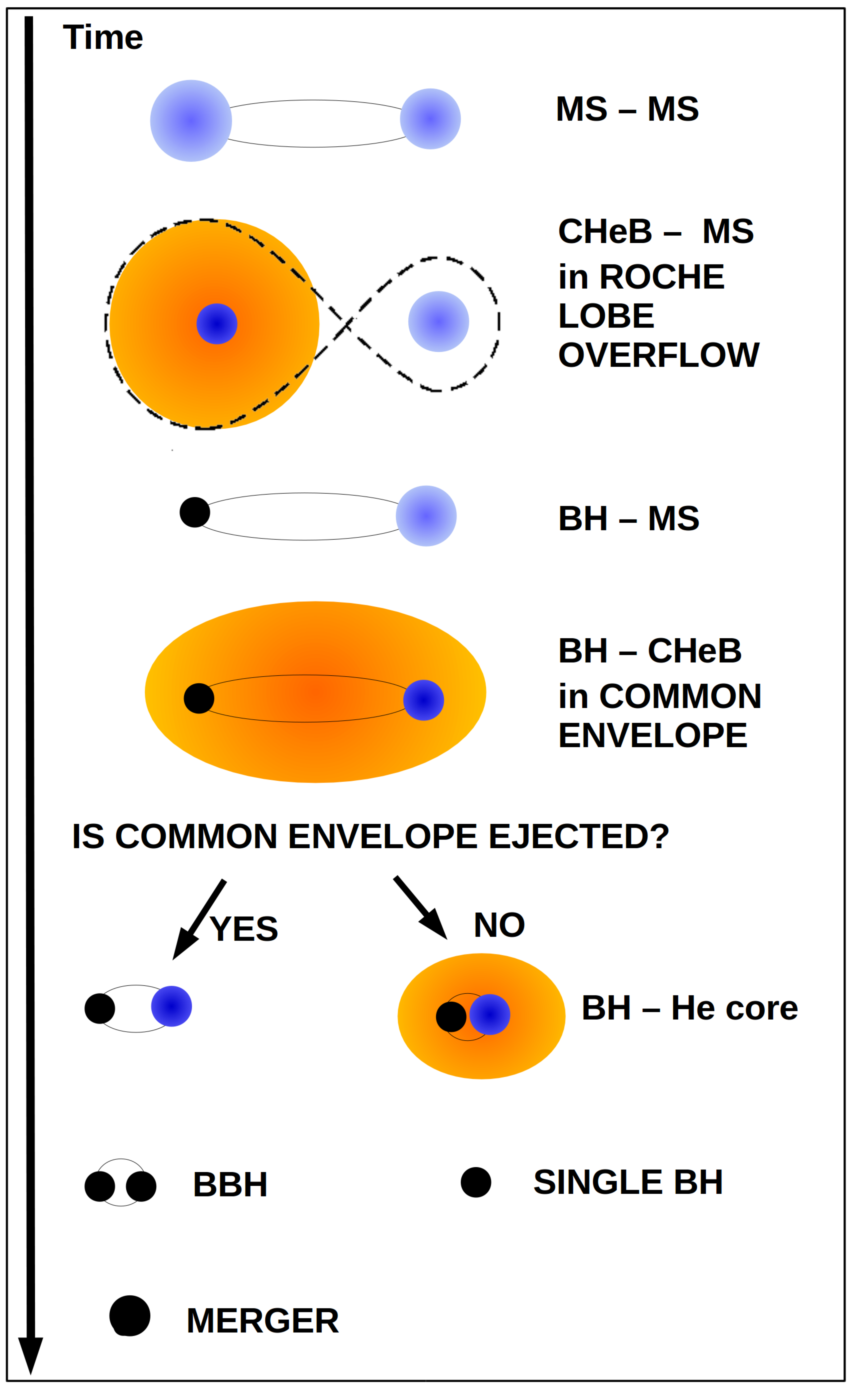}     
}
\caption{\label{fig:cartoonIB}Schematic evolution of an isolated binary star which can give birth to a BBH merger (see e.g. \cite{belczynski2016,mapelli2017,stevenson2017,giacobbo2018a}). }
\end{figure}

To summarize, let us illustrate schematically the evolution of an isolated stellar binary (see e.g. \cite{belczynski2016,mapelli2017,stevenson2017,giacobbo2018a}) which can give birth to merging BHs like GW150914 and the other massive BHs  observed by the LVC \cite{abbottGW150914,abbottGW151226,abbottO1,abbottGW170104,abbottO2,abbottO2popandrate}. In the following discussion, several details of stellar evolution have been simplified or skipped to facilitate the reading for non specialists.

Figure~\ref{fig:cartoonIB} shows the evolution of an isolated binary system composed of two massive stars. These stars are gravitationally bound since their birth. 
Initially, the two stars are both on the main sequence (MS). When the most massive one leaves the MS, 
its radius starts inflating and can grow by a factor of a hundreds. 
The most massive star becomes a giant star with a helium core and a large hydrogen envelope. If its radius equals the Roche lobe (equation~\ref{eq:rlobe}), the system starts a stable mass-transfer episode. Some mass is lost by the system, some is transferred to the companion. After several additional evolutionary stages, the primary collapses to a BH. A direct collapse is preferred with respect to a SN explosion if we want the BH mass to be $>20$ M$_\odot$ (as most BBHs observed by the LVC) and if we want the binary to remain bound (direct collapse implies almost no kick, apart from neutrino mass loss). At this stage, the semi-major axis of the system is still quite large: hundreds to thousands of solar radii.  

When the secondary also leaves the MS, growing in radius, the system enters a CE phase: the BH and the helium core spiral in. If the orbital energy and the thermal energy of the envelope are not sufficient to unbind the envelope, then the BH merges with the helium core leaving a single BH. In contrast, if the envelope is ejected, we are left with a new binary system, composed of the BH and of a stripped naked helium star. The new binary system has a much smaller orbital separation (tens of solar radii) than the pre-CE binary, because of the spiral in. If this new binary system remains bound after the naked helium star undergoes a SN explosion or if the naked helium star is sufficiently massive to directly collapse to a BH, the system evolves into a tight BBH, which might merge within a Hubble time.

The most critical quantities in this scenario are the masses of the two stars and also their initial orbital separation (with respect to the stellar radii): a BBH can merge within a Hubble time only if its initial orbital separation is tremendously small (few tens of solar radii, unless the eccentricity is rather extreme, \cite{peters1964}), but a massive star ($>20$ M$_\odot$) can reach a radius of several thousand solar radii during its evolution. Thus, if the initial orbital separation of the binary star is tens of solar radii, the binary system merges before it can become a BBH. On the other hand, if the initial orbital separation is too large, the two BHs will never merge. In this scenario, the two BHs can merge only if the initial orbital separation of the progenitor binary star is in the range which allows the binary to enter CE and then to leave a short period BBH. This range of initial orbital separations dramatically depends on CE efficiency and on the details of stellar mass and radius evolution. Two possible alternatives to CE are very efficient stable mass transfer \cite{giacobbo2018b,bavera2021b} and CHE \cite{marchant2016,mandel2016}.

\begin{figure}
\center{
\includegraphics[width=12cm]{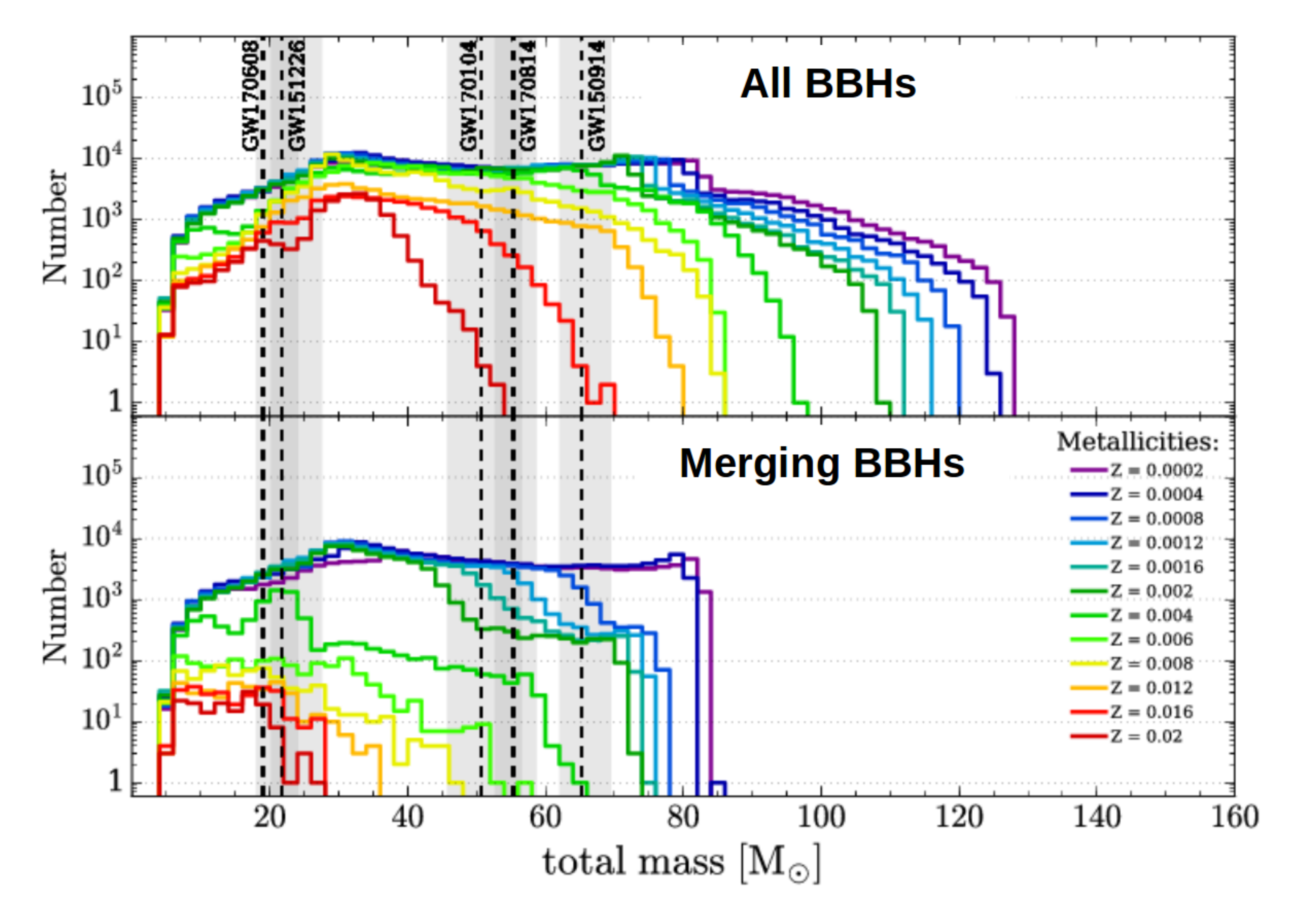}     
}
\caption{\label{fig:giacobbo}Upper panel: total mass ($m_{\rm TOT}=m_1+m_2$) of all BBHs formed from a stellar population of $2\times{}10^7$ massive binary stars simulated with {\sc mobse} and with different metallicities (from 0.0002 to 0.02) \cite{giacobbo2018b}. Lower panel: total mass of the BBHs merging within a Hubble time from the same simulations. 
  See {\url{https://mobse-webpage.netlify.app/}} for more details on {\sc mobse} and on these runs. Courtesy of Nicola Giacobbo.}
\end{figure}

There is one more interesting comment to add. Some stellar wind models predict a maximum BH mass of $\sim{}65-70$ M$_\odot$ from single stellar evolution and from the collapse of the residual hydrogen envelope at solar metallicity (e.g., \cite{giacobbo2018a,giacobbo2018b}). This means that the maximum total mass of a BBH is $m_{\rm TOT}=m_1+m_2\sim{}130-140$ M$_\odot$. However, the maximum total mass of a BBH which merges within a Hubble time is only $m_{\rm TOT}\sim{}80-90$ M$_\odot$ \cite{giacobbo2018b}, as shown in Figure~\ref{fig:giacobbo}. The reason for this difference is that if the initial orbital separation is sufficiently small ($a\sim{}10^2-10^4$ R$_\odot$, \cite{spera2019}) to allow mass transfer and CE, then the two stars lose almost all their hydrogen envelope: they might become a BBH that merges within a Hubble time, but their total mass is at most equal to the total mass of the two naked helium cores. In contrast, if the two stars are metal poor ($Z\lesssim{}0.0002$) and if the initial orbital separation is too large to induce mass transfer and CE, the binary star becomes a BBH almost without mass loss, with a total mass up to $140$ M$_\odot$, but the final semi-major axis is too large to allow coalescence by GW emission.

\section{The dynamics of binary black holes (BBHs)}

In the previous Sections of this review, we discussed the formation of BBHs as isolated binaries. There is an alternative channel for BBH formation: the dynamical evolution scenario.

\subsection{Dynamically active environments}

Collisional dynamics is important for the evolution of binaries only if they are in a dense environment ($\gtrsim{}10^3$ stars pc$^{-3}$), such as a star cluster. On the other hand, astrophysicists believe that the vast majority of massive stars (which are BH progenitors) form in star clusters (\cite{lada2003,weidner2006,weidner2010,portegieszwart2010}, but see also \cite{ward2020}). 

There are several different flavours of star clusters. 
{\emph{Globular clusters}} \cite{gratton2019} are old stellar systems ($\sim{}12$ Gyr), mostly very massive (total mass $M_{\rm SC}\ge{}10^4$ M$_\odot$) and dense (central density $\rho_{\rm c}\ge{}10^4$ M$_\odot$ pc$^{-3}$). They are sites of intense dynamical processes, such as the gravothermal catastrophe. Globular clusters represent a small fraction of the baryonic mass in the local Universe ($\lesssim{}1$ per cent, \cite{harris2013}). Most studies of dynamical formation of BBHs focus on globular clusters (e.g. \cite{sigurdsson1993,portegieszwart2000,mapelli2005,downing2010,downing2011,benacquista2013,tanikawa2013,breen2013MNRAS.432.2779B,samsing2014,rodriguez2015,rodriguez2016,askar2017,samsing2017,samsing2018,askar2018,arcasedda2018MNRAS.479.4652A,fragione2018PhRvL.121p1103F,rodriguez2018,rodriguez2019,kremer2019PhRvD..99f3003K}).

\begin{figure}
\center{
\includegraphics[width=12cm]{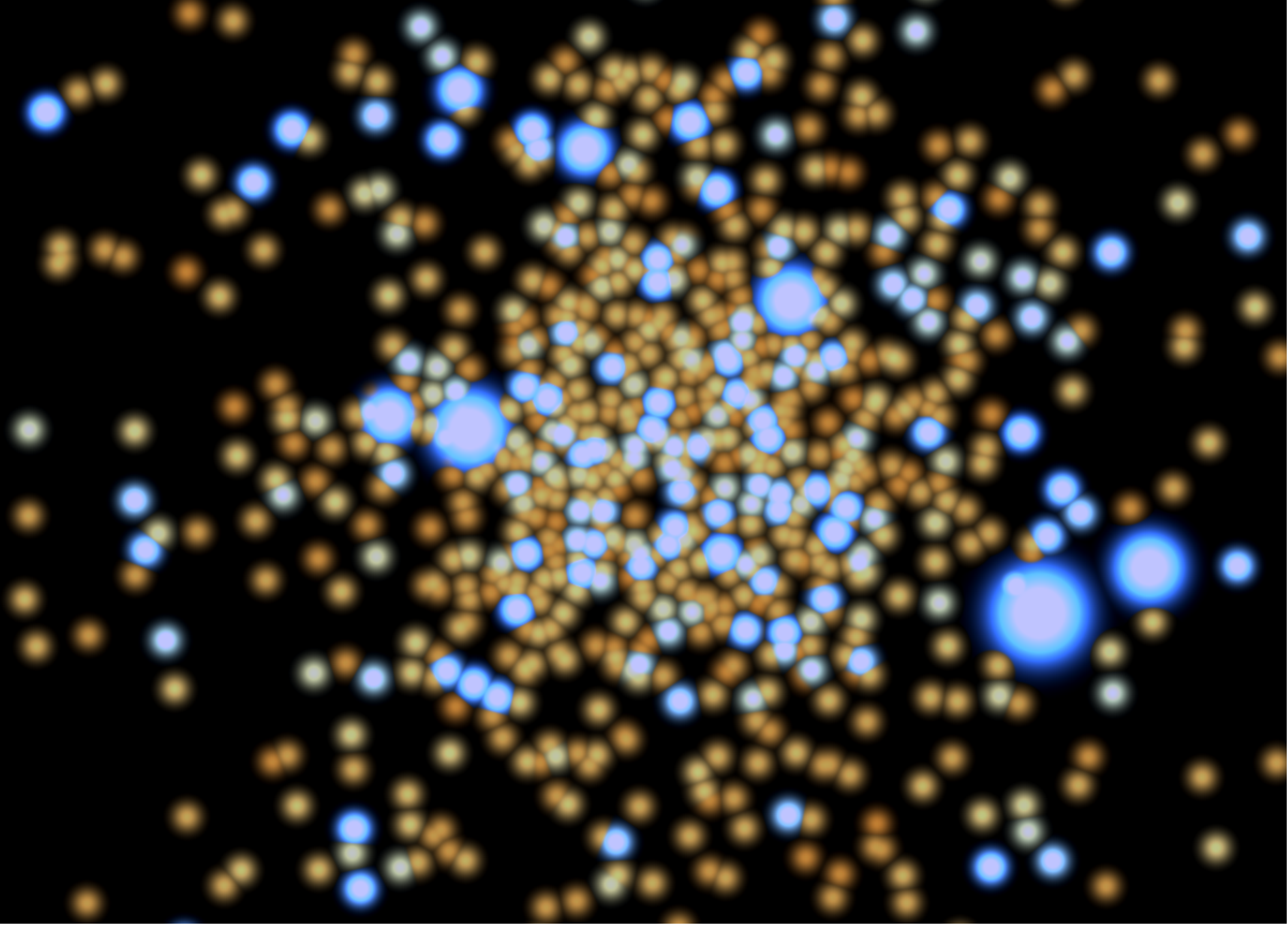}     
}
\caption{\label{fig:starcluster}Snapshot of an N-body simulation of a young star cluster. Different particle sizes and colors refer to different star luminosities and temperatures, as estimated by population-synthesis calculations. This simulation has been included in \cite{mapelli2013}.}
\end{figure}

{\it Young star clusters} are young ($\lesssim{}100$ Myr), relatively dense ($\rho_{\rm c}>10^3$ M$_\odot$ pc$^{-3}$) stellar systems, and are thought to be the most common birthplace of massive stars \cite{lada2003,portegieszwart2010}. When they disperse by gas evaporation or are disrupted by the tidal field of their host galaxy, their stellar content is released into the field. Thus, it is reasonable to expect that a large fraction of BBHs which are now in the field may have formed in young star clusters, where they participated in the dynamics of the cluster. 
A fraction of young star clusters might survive gas evaporation and tidal disruption and evolve into older {\it open clusters}, like M67. Figure~\ref{fig:starcluster} shows a snapshot of an N-body simulation of a young star cluster. Population-synthesis prescriptions are included in this simulation to follow the evolution of binary stars and the formation of BBHs. Studies of BBHs in young and open clusters include \cite{portegieszwart2000,banerjee2010,ziosi2014,mapelli2016,kimpson2016,banerjee2017,banerjee2018,fujii2017,kumamoto2019,kumamoto2020,banerjee2020,dicarlo2019,dicarlo2020a,dicarlo2020b}.

Finally, {\it nuclear star clusters} 
lie in the nuclei of galaxies. Nuclear star clusters are rather common in galaxies (e.g. \cite{boeker2002,ferrarese2006,graham2009,neumayer2020}), are usually more massive and denser than globular clusters, and may co-exist with super-massive BHs (SMBHs). Stellar-mass BHs formed in the innermost regions of a galaxy could even be ``trapped'' in the accretion disc of the central SMBH, triggering their merger (see e.g. \cite{stone2016,bartos2017,mckernan2018}). These features make nuclear star clusters unique among star clusters, for the effects that we will describe in the next Sections.

\subsection{Two-body encounters, dynamical friction and core-collapse}

The main driver of the dynamics of star clusters is gravity force. Gravitational two-body encounters between stars lead to local fluctuations in the potential of the star cluster and drive major changes in the internal structure of the star cluster over a two-body relaxation timescale \cite{spitzer1971,spitzer1987}:
\begin{equation}
 t_{\rm rlx}=0.34\,{}\frac{\sigma^3}{G^2\,{}\langle{}m\rangle{}\,{}\rho{}\,{}\ln{\Lambda}},
\end{equation}
where $\sigma{}$ is the local velocity dispersion of the star cluster, $\langle{}m\rangle{}$ is the average stellar mass in the star cluster, $\rho{}$ is the local mass density, $G$ is the gravity force and $\ln{\Lambda}\sim{}10$ is the Coulomb logarithm. The two-body relaxation timescale is the time needed for a typical star in the stellar system to completely lose memory of its initial velocity due to two-body encounters. In star clusters, $t_{\rm rlx}$ is much shorter than the Hubble time ($t_{\rm rlx}\sim{}10-100$ Myr in young star clusters, \cite{portegieszwart2010}), while in galaxies and large-scale structures it is much longer than the lifetime of the Universe. Hence, close encounters are common in dense star clusters.

Dynamical friction is another consequence of gravity force: a massive body of mass $M$ orbiting in a sea of lighter particles feels a drag force that slows down its motion over a timescale \cite{chandrasekhar1943}:
\begin{equation}\label{eq:dynfric}
  t_{\rm DF}(M)=\frac{3}{4\,{}\left(2\,{}\pi\right)^{1/2}\,{}G^2\,{}\ln{\Lambda}}\,{}\frac{\sigma^3}{M\,{}\rho{}(r)}.
\end{equation}

It is apparent that two-body relaxation and dynamical friction are driven by the same force and are related by 
\begin{equation}
  t_{\rm DF}(M)\sim{}\frac{\langle{}m\rangle{}}{M}\,{}t_{\rm rlx},
\end{equation}
i.e., dynamical friction happens over a much shorter timescale than two-body relaxation and leads to mass segregation (or mass stratification) in a star cluster. This process speeds up the collapse of the core of a star cluster and can trigger the so-called Spitzer's instability \cite{spitzer1969}. 
Two-body relaxation, dynamical friction and their effects play a crucial role in shaping the demography of BBHs in star clusters, as we discuss below.

\subsection{Binary -- single encounters}

We now review what are the main dynamical effects which can affect a BBH, starting from binary -- single star encounters. Binaries have a energy reservoir, their internal energy:
\begin{equation}
E_{\rm int}=\frac{1}{2}\,{}\mu{}\,{}v^2-\frac{G\,{}m_1\,{}m_2}{r},
\end{equation}
where $\mu{}=m_1\,{}m_2/(m_1+m_2)$ is the reduced mass of the binary (whose components have mass $m_1$ and $m_2$), $v$ is the relative velocity between the two members of the binary, and $r$ is the distance between the two members of the binary. As shown by Kepler's laws, $E_{\rm int}=-E_{\rm b}=-G\,{}m_1\,{}m_2/(2\,{}a)$, where $E_{\rm b}$ is the binding energy of the binary system, $a$ being its semi-major axis.

The internal energy of a binary can be exchanged with other stars only if the binary undergoes a close encounter with a star, so that its orbital parameters are perturbed by the intruder. This happens only if a single star approaches the binary by few times its orbital separation. We define this close encounter between a binary and a single star as a {\it three-body encounter}. For this to happen with a non-negligible frequency, the binary must be in a dense environment, because the rate of three-body encounters scales with the local density of stars. Three-body encounters have crucial effects on BH binaries, such as {\it hardening}, {\it exchanges}  and {\it ejections}.

\subsection{Hardening}

If a BBH undergoes a number of three-body encounters during its life, we expect that its semi-major axis will shrink as an effect of the encounters. This process is called dynamical {\it hardening}.

Following \cite{heggie1975}, we call hard binaries (soft binaries) those binaries with binding energy larger (smaller) than the average kinetic energy of a star in the star cluster. According to Heggie's law \cite{heggie1975}, hard binaries tend to harden (i.e., to become more and more bound) via binary--single encounters. In other words, a fraction of the internal energy of a hard binary can be transferred into kinetic energy of the intruder and of the centre-of-mass of the binary during three body encounters. This means that the binary loses internal energy and its semi-major axis shrinks. 

Most BBHs are expected to be hard binaries, because BHs are among the most massive bodies in star clusters. Thus, BBHs are expected to harden as a consequence of three-body encounters. The hardening process may be sufficiently effective to shrink a BBH till it enters the regime where GW emission is efficient: a BBH which is initially too loose to merge may then become a GW source thanks to dynamical hardening. The hardening rate for hard binaries with semi-major axis $a$ can be estimated as \cite{heggie1975}
\begin{equation}\label{eq:hardening}
  \frac{{\rm d}}{{\rm d}t}\left(\frac{1}{a}\right)=2\,{}\pi{}\,{}G\,{}\xi{}\,{}\frac{\rho{}}{\sigma{}},
\end{equation}
where $\xi{}\sim{}0.1-10$ is a dimensionless hardening parameter (which has been estimated through numerical experiments, \cite{hills1983,quinlan1996}), $\rho{}$ is the local mass density of stars, $\sigma{}$ is the local velocity dispersion and $G$ is the gravity constant.

Dynamical hardening is the main responsible for the shrinking of a binary, till it reaches a semi-major axis short enough for GW emission to become effective, which can be derived with the following equation \cite{peters1964}:
\begin{equation}\label{eq:peters}
  \frac{{\rm d}a}{{\rm d}t}=-\frac{64}{5}\frac{G^3\,{}m_1\,{}m_2\,{}(m_1+m_2)}{c^5\,{}(1-e^2)^{7/2}}\,{}a^{-3}.
\end{equation}

By combining equations~\ref{eq:hardening} and \ref{eq:peters}, it is possible to make a simple analytic estimate of the evolution of the semi-major axis of a BBH  which is affected by three-body encounters and by GW emission: 
\begin{equation}\label{eq:miatesi}
\frac{da}{dt}=-2\,{}\pi{}\,{}\xi{}\,{}\frac{G\,{}\rho{}}{\sigma{}}\,{}a^2-\frac{64}{5}\frac{G^3\,{}m_1\,{}m_2\,{}(m_1+m_2)}{c^5\,{}(1-e^2)^{7/2}}\,{}a^{-3},
\end{equation}
This equation holds under the assumption that the binary star is hard, the total mass of the binary star is much larger than the average mass of a star in the star cluster (exchanges are unlikely) and that most three-body encounters have a small impact parameter. The first part of the right-hand term of equation~\ref{eq:miatesi} accounts for the effect of three-body hardening on the semi-major axis. It scales as $da/dt\propto{}-a^2$, indicating that the larger the binary is, the more effective the hardening. This can be easily understood considering that the geometric cross section for three body interactions with a binary scales as $a^2$. The second part of the right-hand term of equation~\ref{eq:miatesi} accounts for energy loss by GW emission. It is the first-order approximation of the calculation by Peters (1964, \cite{peters1964}). It scales as $da/dt\propto{}-a^{-3}$ indicating that GW emission becomes efficient only when the two BHs are very close to each other.

\begin{figure}
\center{
\includegraphics[width=12cm]{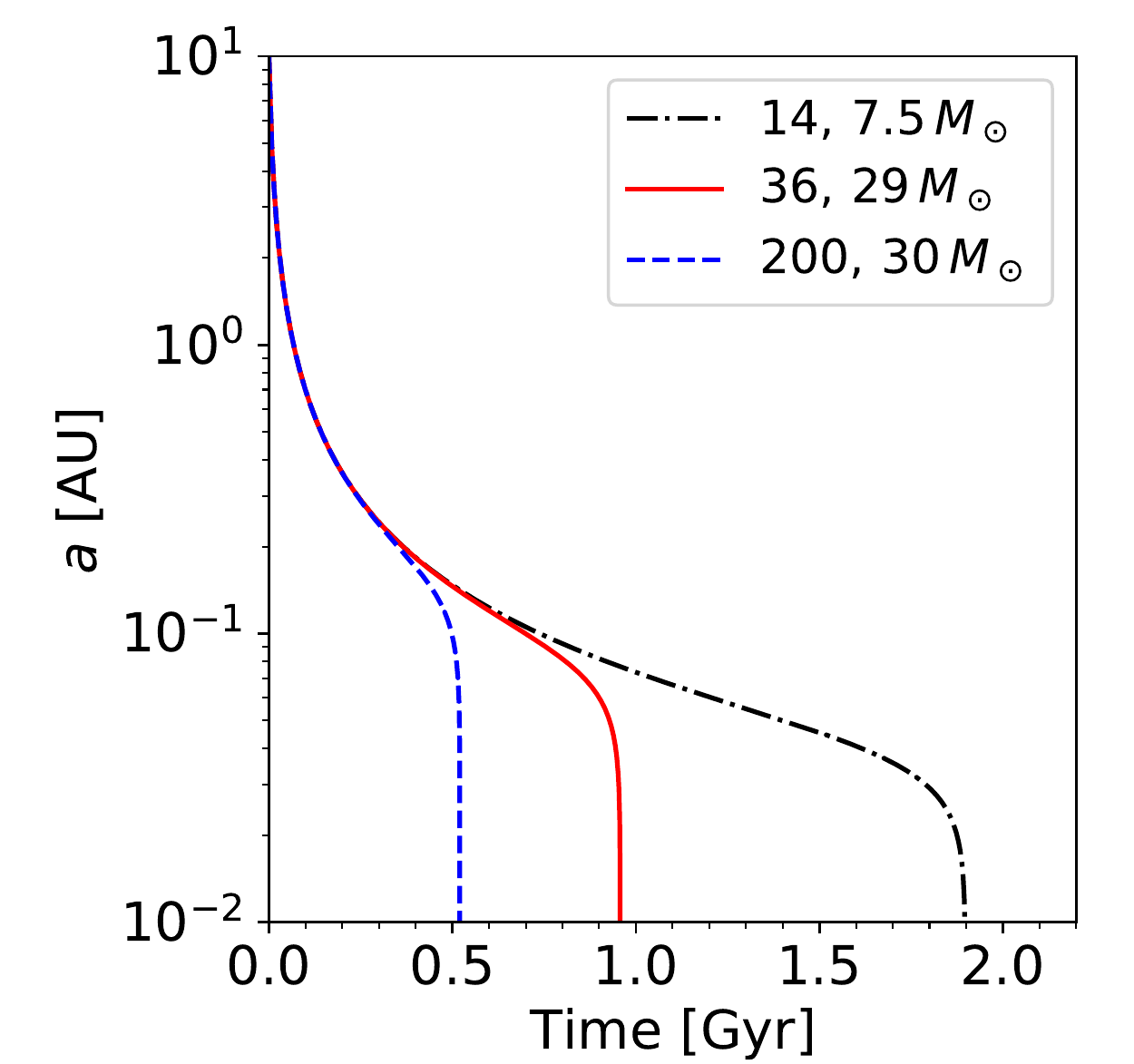}     
}
\caption{\label{fig:miatesi}Time evolution of the semi-major axis of three BH binaries estimated from equation~\ref{eq:miatesi}. Blue dashed line: BH binary with masses $m_1=200$ M$_\odot$, $m_2=30$ M$_\odot$; red solid line: $m_1=36$ M$_\odot$, $m_2=29$ M$_\odot$; black dot-dashed line: $m_1=14$ M$_\odot$, $m_2=7.5$ M$_\odot$. For all BH binaries: $\xi{}=1$, $\rho{}=10^5$ M$_\odot$ pc$^{-3}$, $\sigma{}=10$ km s$^{-1}$, $e=0$ (here we assume that $\rho{}$, $\sigma{}$ and $e$ do not change during the evolution), initial semi-major axis of the BH binary $a_{\rm i}=10$ AU.}
\end{figure}

In Figure~\ref{fig:miatesi} we solve equation~\ref{eq:miatesi} numerically for three BBHs with different mass. All binaries evolve through i) a first phase in which hardening by three body encounters dominates the evolution of the binary, ii) a second phase in which the semi-major axis stalls because three-body encounters become less efficient as the semi-major axis shrink, but the binary is still too large for GW emission to become efficient, and iii) a third phase in which the semi-major axis drops because the binary enters the regime where GW emission is efficient.

\subsection{Exchanges}

Dynamical exchanges are three-body encounters during which one of the members of the binary is replaced by the intruder. 
Exchanges may lead to the formation of new BBHs:  
if a binary composed of a BH and a low-mass star undergoes an exchange with a single BH, this leads to the formation of a new BBH. This is a fundamental difference between BHs in the field and in star clusters: a BH which forms as a single object in the field has negligible chances to become member of a binary system, while a single BH in the core of a star cluster has good chances of becoming member of a binary by exchanges.

Exchanges are expected to lead to the formation of many more BBHs than they can destroy, because the probability for an intruder to replace one of the members of a binary is $\approx{}0$ if the intruder is less massive than both binary members, while it suddenly jumps to $\sim{}1$ if the intruder is more massive than one of the members of the binary \cite{hills1980}. Since BHs are among the most massive bodies in a star cluster 
after their formation, they are very efficient in acquiring companions through dynamical exchanges. Thus, exchanges are a crucial mechanism to form BH binaries dynamically. By means of direct N-body simulations, Ziosi et al. (2014, \cite{ziosi2014}) show that $>90$ per cent of BBHs in young star clusters form by dynamical exchange. Moreover, BBHs formed via dynamical exchange will have some distinctive features with respect to field BBHs (see e.g. \cite{ziosi2014}):
\begin{itemize}
\item{}BBHs formed by exchanges will be (on average) more massive than isolated BBHs, because more massive intruders have higher chances to acquire companions;
\item{}exchanges trigger the formation of highly eccentric BBHs; eccentricity is then significantly reduced by circularisation induced by GW emission, if the binary enters the regime where GW emission is effective;
\item{}BBHs born by exchange will likely have misaligned spins: exchanges and other dynamical interactions tend to lead to isotropically distributed spin directions with respect to the binary orbital plane, because dynamical interactions remove any memory of previous alignments.
\end{itemize}


Zevin et al. (2017, \cite{zevin2017}) compare a set of simulations of field binaries with a set of simulations of globular cluster binaries, run with the same population-synthesis code. The most striking difference between merging BH binaries in their globular cluster simulations and in their population-synthesis simulations is the dearth of merging BHs with mass $<10$ M$_\odot$ in the globular cluster simulations. This is likely due to the fact that exchanges tend to destroy binaries composed of light BHs.

\begin{figure}
\center{
\includegraphics[width=12cm]{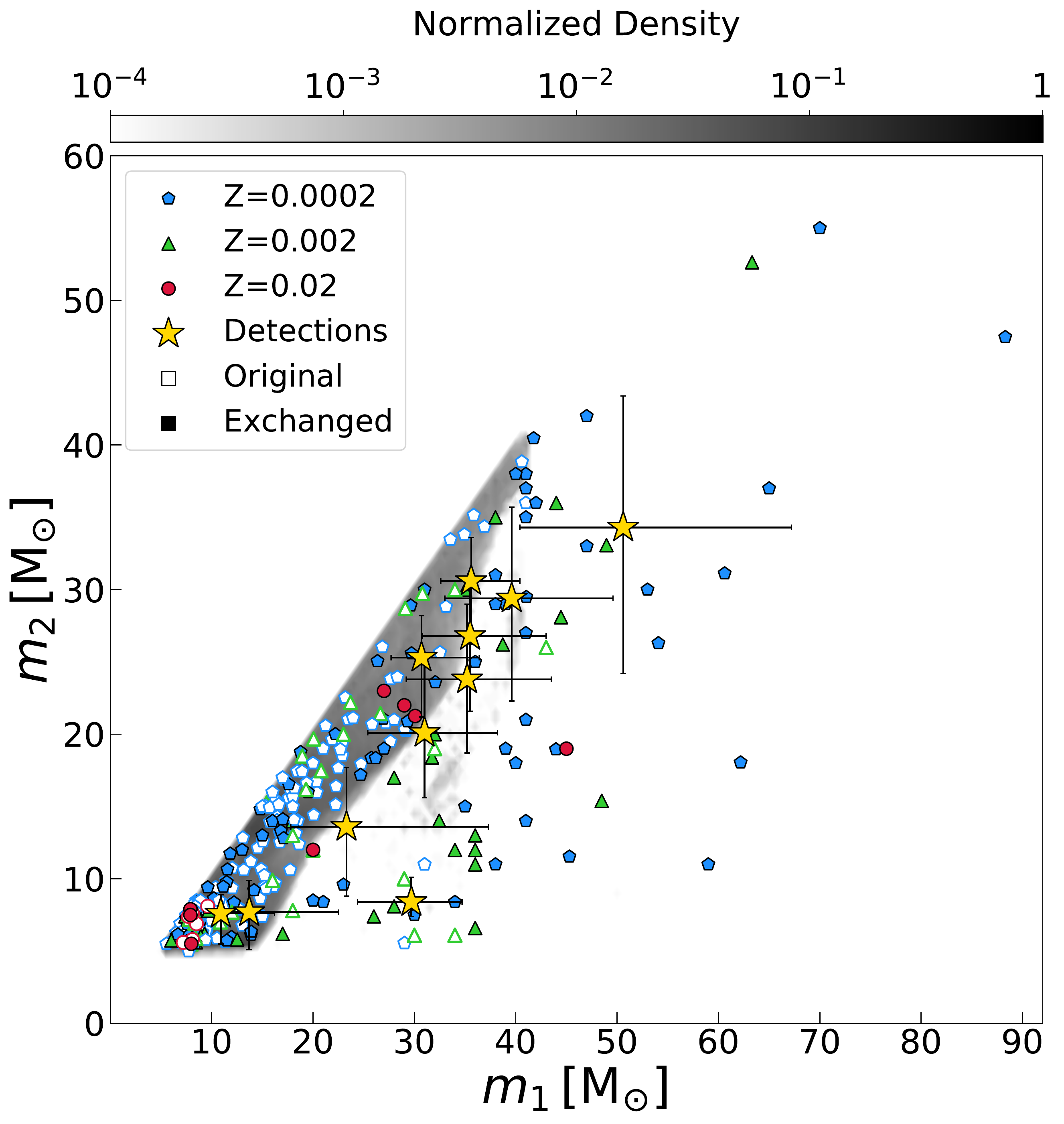}     
}
\caption{\label{fig:ugo} Mass of the secondary BH versus mass of the primary BH in a set of BBH mergers. Yellow stars: O1 and O2 BBHs plus GW190412. Blue, green and red symbols: simulated BBHs in young star clusters with metallicity $Z=0.0002,$ 0.002 and 0.02. Filled symbols: dynamical exchanges. Open symbols: original binaries. Gray contours: BBHs from isolated binary evolution. Simulations from \cite{dicarlo2020b}. Courtesy of Ugo N. Di Carlo.}
\end{figure}

Figure~\ref{fig:ugo} compares the masses of simulated BBH mergers in isolated binaries and in young star clusters. While the maximum total mass of BBH mergers in isolated binaries is $\sim{}80$ M$_\odot$ (see Figure~\ref{fig:giacobbo}), BBH mergers in young star clusters have total masses up to $\sim{}130$ M$_\odot$. This difference springs mainly from two facts:
\begin{itemize}
  \item[i)] single stellar evolution at low metallicity can lead to the formation of single BHs with mass up to $\sim{}70$ M$_\odot$, if the hydrogen envelope collapses to the final BH. In the field such massive BHs remain single, while in a star cluster they can acquire companions via dynamical exchanges and they can merge by dynamical hardening and GW emission;
  \item[ii)] in star clusters, stellar collisions are quite common and can lead to the build up of more massive BHs, even with mass inside the pair instability mass gap \cite{dicarlo2019,dicarlo2020a,dicarlo2020b,kremer2020,renzo2020}.
\end{itemize}

Spin misalignments are another possible feature to discriminate between field binaries and star cluster binaries (e.g. \cite{farr2017,farr2018}). 
We expect that an isolated binary system in which both the primary and the secondary component undergo direct collapse results in a BBH with nearly aligned spins. For dynamically formed BH binaries we expect misaligned, or even nearly isotropic spins, because any original spin alignment is completely reset by three-body encounters.

Currently, we have limited constraints on BH spins from GW detections.  In a few events, such as GW151226 \cite{abbottGW151226}, GW170729 \cite{abbottO2} and especially GW190412 \cite{abbottGW190412}, the measured value of $\chi_{\rm eff}$ is significantly larger than zero, indicating at least partial alignment. 
The vast majority of  events listed in GWTC-2 have $\chi_{\rm eff}$ consistent with zero. This might be the result of either low spins or misaligned spins or a combination of both.  On the other hand, the most recent population study by the LVC \cite{abbottO3apopandrate} shows that  $\sim{}12$\% to 44\% of BBH systems have spins tilted by more than 90$^\circ$ with respect to their orbital angular momentum, supporting a negative effective spin parameter.

Interestingly, GW190412 has a non-zero precessing spin $\chi_{\rm p}$ to the 90\% credible level \cite{abbottGW190412} and GW190521 \cite{abbottGW190521astro}, the most massive BBH event to date,  shows mild evidence for non-zero precessing spin. This might support a dynamical formation for these events (e.g., \cite{abbottGW190412,abbottGW190521astro,romeroshaw2020arXiv200904771R,gayathri2020arXiv200905461G,fragione2020arXiv200905065F}).



\subsection{Stellar mergers and BHs in the pair-instability mass gap}

\begin{figure}
\center{
\includegraphics[width=12cm]{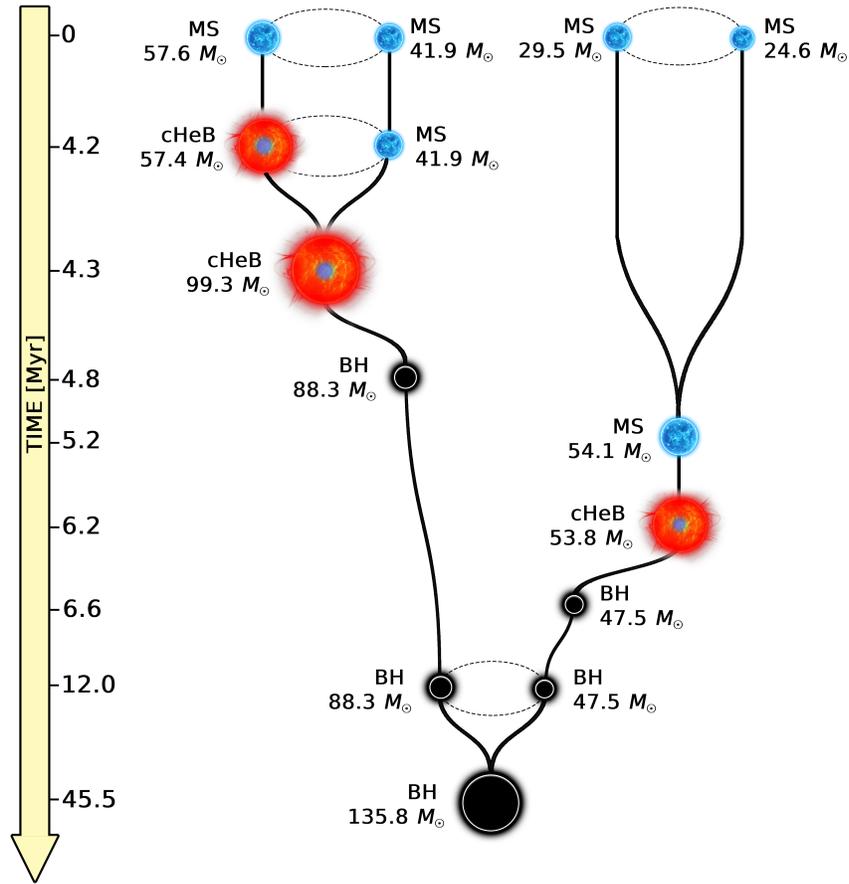}     
}
\caption{\label{fig:gw190521} Cartoon of the dynamical assembly of a GW190521-like BBH from the simulations of \cite{dicarlo2020b}. Courtesy of Ugo N. Di Carlo.}
\end{figure}
Binary--single encounters in young star clusters trigger collisions between massive stars. This process is sped up by dynamical friction, which can generate an excess of massive stars in the cluster core in less than a Myr. Stellar collisions can lead to the formation of very massive stars ($>150$ M$_\odot$), blue straggler stars \cite{mapelli2006} and unusually massive BHs.

Figure~\ref{fig:gw190521} shows the dynamical formation of a BBH with primary mass $m_1=88$ M$_\odot$ and secondary mass $m_2=47.5$ M$_\odot$ \cite{dicarlo2020b}. The masses of the components of this BBH are similar to those of GW190521 \cite{abbottGW190521,abbottGW190521astro}. In particular, the primary BH has a mass in the pair-instability mass gap. Its formation is possible because its stellar progenitor is the result of the merger between a giant star with a well-developed He core and a main sequence companion (MS). The result of this stellar merger is a massive core helium burning (cHeB) star with an over-sized hydrogen envelope with respect to the He core \cite{dicarlo2019,dicarlo2020a,dicarlo2020b,kremer2020,renzo2020}. Given the short timescale for He, C, O, Ne and Si burning with respect to H burning, the star collapses to a BH before the He core grows above the threshold for pair instability: the result is a $\sim{}88$ M$_\odot$ BH. After its formation, this BH acquires a companion by dynamical exchanges and merges within a Hubble time.

This is a viable scenario to produce systems like GW190521 not only because of the good match of the BBH mass, but also because a BH born from the direct collapse of a very massive star might form with a large spin (depending on the final spin of the massive progenitor) and because this dynamical formation leads to isotropically oriented spins. The main uncertainties of this model are mass loss during stellar collision \citep{gaburov2010} and the possible loss of a fraction of the envelope during the final collapse \citep{fernandezquataert2018}.

\subsection{Direct three-body binary formation}

In the most massive star clusters (globular clusters, nuclear star clusters), stellar velocities are so large that dynamical encounters can unbind most of the original binary stars (i.e., those binary stars that were already there at the formation of the star cluster). The minimum relative velocity $v_{\rm c}$ between a binary star and an intruder star needed to unbind a binary is in fact \cite{sigurdsson1995}:
\begin{equation}
  v_{\rm c}=\sqrt{\frac{G\,{}m_1\,{}m_2\,{}(m_1+m_2+m_3)}{m_3\,{}(m_1+m_2)\,a}},
\end{equation}
where $m_1$, $m_2$ and $m_3$ are the masses of the two binary members and that of the intruder, while $a$ is the binary semi-major axis.

In these extreme environments, most BBHs are expected to form by direct encounters of three single bodies \cite{morscher2015,samsing2017}, during core collapse. This leads to the formation of extremely hard BBHs, which survive further ionization from intruders. The timescale for binary formation via three single body encounters is \cite{lee1995,antonini2016}:
\begin{equation}\label{eq:t3bb}
t_{\rm 3bb}=0.1\,{}{\rm Myr}\,{}\left(\frac{n}{10^6\,{}{\rm pc}^{-3}}\right)^{-2}\,{}\left(\frac{\sigma{}}{30\,{}{\rm km}\,{}{\rm s}^{-1}}\right)^9\,{}\left(\frac{m_{\rm BH}}{30\,{}M_\odot}\right)^{-5},
\end{equation}
where $n$ is the local stellar density, $\sigma{}$ is the local velocity dispersion of the star cluster and $m_{\rm BH}$ is the typical BH mass in the star cluster. 
The properties of BBHs born from three single body encounters are similar to those of BBHs born via dynamical exchanges: they tend to be more massive than isolated binaries, have high initial eccentricity and isotropically oriented spins (e.g. \cite{antonini2020}).

Direct three-body encounters are likely the most common BBH formation channel in globular clusters and nuclear star clusters \cite{morscher2015}, while binary -- single star exchanges are likely the most common formation channel of BBHs in young star clusters \cite{ziosi2014,dicarlo2019}. In young star clusters, dynamical exchanges affect both already formed BHs and their stellar progenitors.

\subsection{Dynamical ejections}

During three-body encounters, a fraction of the internal energy of a hard binary is transferred into kinetic energy of the intruders and of the centre-of-mass of the binary. As a consequence, the binary and the intruder recoil. The recoil velocity is generally of the order of a few km s$^{-1}$, but can be up to several hundred km s$^{-1}$.

Since the escape velocity from a globular cluster is $\sim{}30$ km s$^{-1}$ and the escape velocity from a young star cluster or an open cluster is even lower, both the recoiling binary and the intruder can be ejected from the parent star cluster. If the binary and/or the intruder are ejected, they become field objects and cannot participate in the dynamics of the star cluster anymore. Thus, not only the ejected BBH stops hardening, but also the ejected intruder, if it is another compact object, loses any chance of entering a new binary by dynamical exchange.

A general expression for the recoil velocity of the binary center  of mass if $(m_1+m_2)\gg{}\langle{}m\rangle{}$ (where $\langle{}m\rangle$ is the average mass of a star in a star cluster) is
\begin{equation}
  v_{\rm rec}\sim{}\frac{\langle{}m\rangle{}}{m_1+m_2}\,{}\sqrt{\frac{2\,{}\xi{}}{(m_1+m_2+\langle{}m\rangle{})}\,{}E_{\rm b}},
\end{equation}
where $E_{\rm b}=G\,{}m_1\,{}m_2/(2\,{}a)$ is the binary binding energy. The above equation can help us deriving the minimum binding energy above which a binary star is ejected by a binary--single encounter $E_{\rm b,\,{}min}$ \cite{miller2002}:
\begin{equation}\label{eq:ejec}
  E_{\rm b,\,{}min}\sim{}\frac{(m_1+m_2)^3}{2\,{}\xi{}\,{}\langle{}m\rangle{}^2}\,{}v_{\rm esc}^2,
\end{equation}
where $v_{\rm esc}$ is the escape velocity from the star cluster. A BBH will merge inside the star cluster only if $E_{\rm b,\,{}min}>E_{\rm b,\,{}GW}$, where $E_{\rm b,\,{}GW}$ is the minimum binding energy to reach coalescence by GW emission.


Most BNSs, BBHs and BH--NS systems in young star clusters are ejected before they merge \cite{dicarlo2019,rastello2020}. Dynamical ejections of BNS and BH--NS binaries were proposed to be one of the possible explanations for the host-less short gamma-ray bursts, i.e. gamma-ray bursts whose position in the sky appears to be outside any observed galaxy \cite{fong2013}. Host-less bursts may be $\sim{}25$\% all short gamma-ray bursts.

In general, ejections of compact objects and compact-object binaries from their parent star cluster can be the result of at least three different processes:
\begin{itemize}
\item{}dynamical ejections (as described above);
\item{}SN kicks \cite{hobbs2005,fryer2012};
\item{}GW recoil \cite{lousto2009,campanelli2007,gonzalez2007}.
\end{itemize}

GW recoil is a relativistic kick occurring when a BBH merges. It is the result of asymmetric linear momentum loss by GW emission, when the binary has asymmetric component masses and/or misaligned spins. It results in kick velocities up to thousands of km s$^{-1}$ and usually of the order of hundreds of km s$^{-1}$.

Ejections by dynamics, SN kicks or GW recoil may be the main process at work against mergers of second-generation BHs, where for second-generation BHs we mean BHs which were born from the merger of two BHs rather than from the collapse of a star \cite{gerosa2017}. In globular clusters, open clusters and young star clusters, a BH has good chances of being ejected by three-body encounters before it merges \cite{miller2002} and a very high chance of being ejected by GW recoil after it merges \cite{rodriguez2019,mapelli2020b}. The only place where merging BHs can easily avoid ejection by GW recoil are nuclear star clusters, whose escape velocity can reach hundreds of km s$^{-1}$ \cite{antonini2016,antonini2019,fragione2020,mapelli2020b}.


\subsection{Formation of intermediate-mass black holes by runaway collisions}

In Section~\ref{sec:remnants}, we have mentioned that intermediate-mass black holes (IMBHs, i.e. BHs with mass $100\lesssim{}m_{\rm BH}\lesssim{}10^4$ M$_\odot$) might form from the direct collapse of metal-poor extremely massive stars \cite{spera2017}. Other formation channels have been proposed for IMBHs and most of them involve the dynamics of star clusters. The formation of massive BHs by runaway collisions has been originally proposed about half a century ago \cite{colgate1967,sanders1970} and was then elaborated by several authors (e.g. \cite{portegieszwart1999,portegieszwart2002,portegieszwart2004,gurkan2006,freitag2006,mapelli2006,mapelli2008,giersz2015,mapelli2016,dicarlo2019,dicarlo2020b,rizzuto2020}). 

\begin{figure}
\center{
\includegraphics[width=12cm]{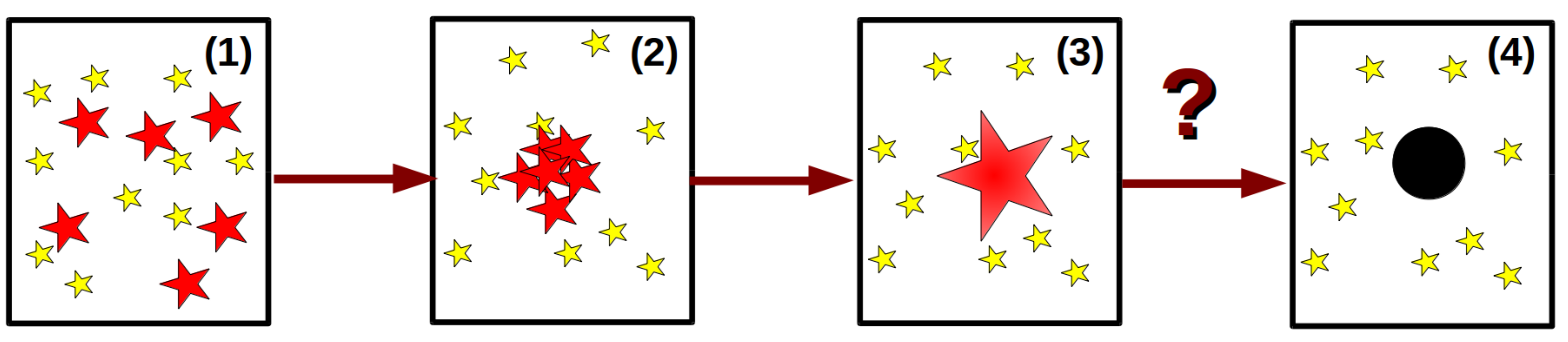}     
}
\caption{\label{fig:runaway}Cartoon of the runaway collision scenario in dense young star clusters (see e.g. \cite{portegieszwart2002}). From left to right: (1) the massive stars (red big stars) and the low-mass stars (yellow small stars) follow the same initial spatial distribution; (2) dynamical friction leads the massive stars to sink to the core of the cluster, where they start colliding between each other; (3) a very massive star ($\gg{}100$ M$_\odot$) forms as a consequence of the runaway collisions; (4) this massive star might be able to directly collapse into a BH.}
\end{figure}

The basic idea is the following (as summarized by the cartoon in Figure~\ref{fig:runaway}). In a dense star cluster, dynamical friction \cite{chandrasekhar1943} makes massive stars to decelerate because of the drag exerted by lighter bodies, on a timescale $t_{\rm DF}(M)\sim{}(\langle{}m\rangle{}/M)\,{}t_{\rm rlx}$ (equation~\ref{eq:dynfric}).

Since the two-body relaxation timescale in a young star cluster can be as short as $t_{\rm rlx}\sim{}10-100$ M$_\odot$ \cite{portegieszwart2010}, for a star with mass $M\ge{}40$ M$_\odot$ we estimate $t_{\rm DF}\le{}2.5$ Myr: dynamical friction is very effective in dense massive young star clusters. Because of dynamical friction, massive stars segregate to the core of the cluster before they become BHs.

If the most massive stars in a dense young star cluster sink to the centre of the cluster by dynamical friction on a time shorter than their lifetime (i.e., before core-collapse SNe take place, removing a large fraction of their mass), then the density of massive stars in the cluster core becomes extremely high. This makes collisions between massive stars extremely likely. Actually, direct N-body simulations show that collisions between massive stars proceed in a runaway sense, leading to the formation of a very massive ($\gg{}100$ M$_\odot$) star \cite{portegieszwart2002}. The main open question is: ``What is the final mass of the collision product? Is the collision product going to collapse to an IMBH?''.


There are essentially two critical issues: i) how much mass is lost during the collisions? ii) how much mass does the very-massive star lose by stellar winds?  

Hydrodynamical simulations of colliding stars \cite{gaburov2008,gaburov2010} show that massive star can lose $\approx{}25$\% of their mass during collisions. Even if we optimistically assume that no mass is lost during and immediately after the collision (when the collision product relaxes to a new equilibrium), the resulting very massive star will be strongly radiation-pressure dominated and is expected to lose a significant fraction of its mass by stellar winds. Recent studies including the effect of the Eddington factor on mass loss \cite{mapelli2016,spera2017} show that IMBHs cannot form from runaway collisions at solar metallicity. At lower metallicity ($Z\lesssim{}0.1$ Z$_\odot$) approximately $10-30$\% of runaway collision products in young dense star clusters can become IMBHs by direct collapse (they also avoid being disrupted by pair-instability SNe).

The majority of runaway collision products do not become IMBHs but they end up as relatively massive BHs ($\sim{}20-90$ M$_\odot$, \cite{mapelli2016}). If they remain inside their parent star cluster, such massive BHs are extremely efficient in acquiring companions by dynamical exchanges. Mapelli (2016, \cite{mapelli2016}) find that all stable binaries formed by the runaway collision product are BBHs and thus are possibly important sources of GWs in the LIGO--Virgo range.

\subsection{Hierarchical BBH formation and IMBHs}

The runaway collision scenario occurs only in the early stages of the evolution of a star cluster, when massive stars are still alive (the lifetime of a $\sim{}30$ M$_\odot$ star is $\sim{}6$ Myr). However, it has been proposed that IMBHs form even in old clusters (e.g., globular clusters) by repeated mergers of smaller BHs (e.g. \cite{miller2002,giersz2015,antonini2019,fragione2020,mapelli2020b,mapelli2020c}).

\begin{figure}
\center{
\includegraphics[width=12cm]{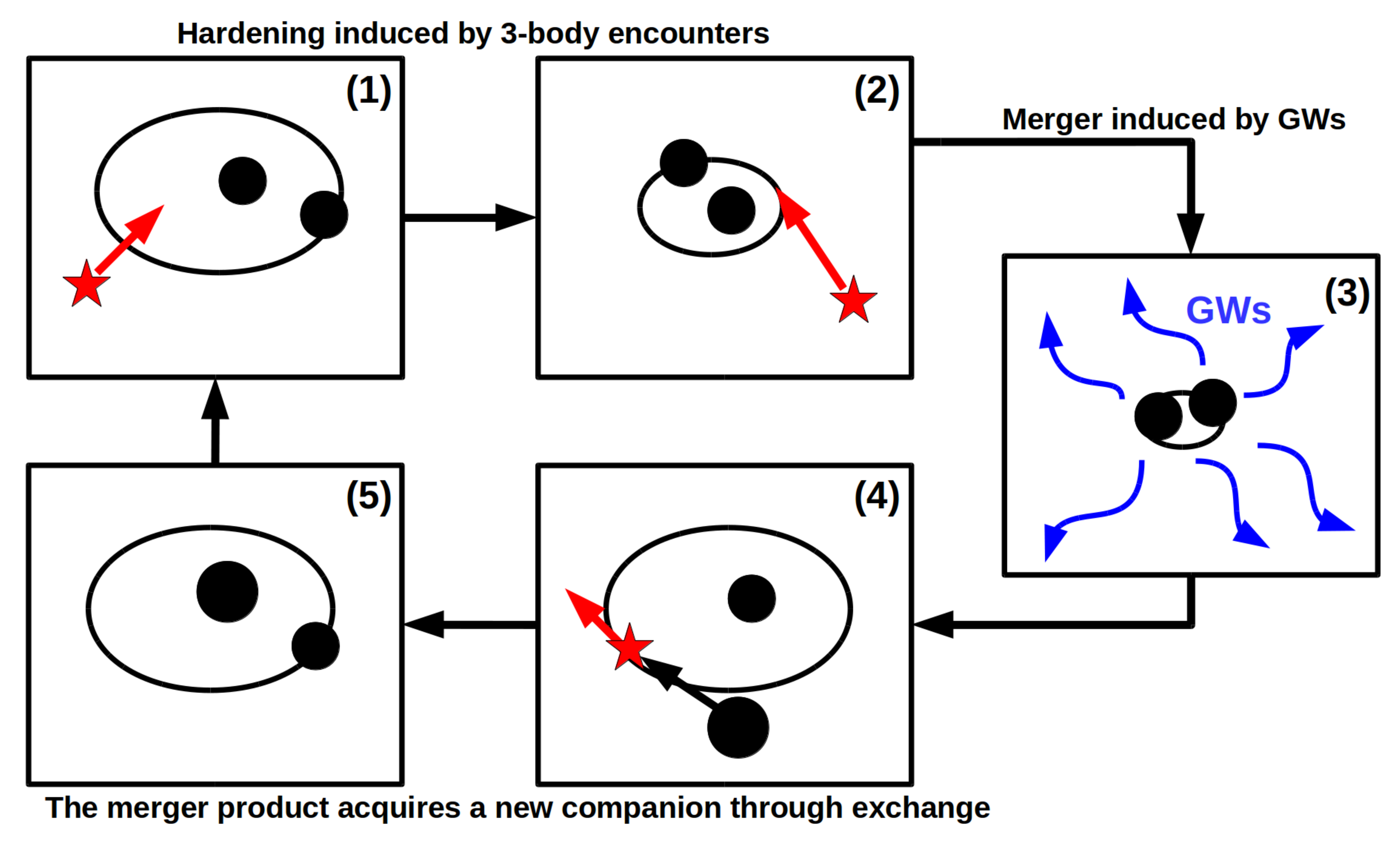}     
}
\caption{\label{fig:repeatedmergers}Cartoon of the repeated merger scenario in old star clusters (see, e.g., \cite{miller2002,giersz2015}). From top to bottom and from left to right: (1) a BBH undergoes three-body encounters in a star cluster; (2) three-body encounters harden the BBH, shrinking its semi-major axis; (3) the BBH hardens by three-body encounters till it enters the regime where GW emission is efficient: the BBH semi-major axis decays by GW emission and the binary merges; (4) a single bigger BH forms as result of the merger and may acquire a new companion by dynamical exchange (if it is not ejected by GW recoil); (5) the new BBH containing the bigger BH starts shrinking again by three-body encounters (1). This loop may be repeated several times till the main BH becomes an IMBH.
}
\end{figure}

The simple idea is illustrated in Figure~\ref{fig:repeatedmergers}. A stellar BBH  in a star cluster is usually a rather hard binary. Thus, it shrinks by dynamical hardening till it may enter the regime where GW emission is effective. In this case, the BBH merges leaving a single more massive BH. Given its relatively large mass, the new BH has good chances to acquire a new companion by exchange. Then, the new BBH starts hardening again by three-body encounters and the story may repeat several times, till the main BH becomes an IMBH.

This scenario has one big advantage: it does not depend on stellar evolution, so we are confident that the BH will grow in mass by mergers, if it remains inside the cluster. However, there are several issues. First, the BBH may be ejected by dynamical recoil, received as an effect of three-body encounters. Recoils get stronger and stronger, as the orbital separation decreases (equation~\ref{eq:ejec}). The BBH will avoid ejection by dynamical recoil only if it is sufficiently massive ($\gtrsim{}50$ M$_\odot$ for a dense globular cluster, \cite{colpi2003}). If the BBH is ejected, the loop breaks and no IMBH is formed.

Second and even more important, the merger of two BHs involves a relativistic kick. This kick may be as large as hundreds of km s$^{-1}$ \cite{lousto2009}, leading to the ejection of the BH from the parent star cluster \cite{holleybockelmann2008}. Also in this case, the loop breaks and no IMBH is formed. Finally, even if the BH binary is not ejected, this scenario is relatively inefficient: if the seed BH is $\sim{}50$ M$_\odot$, several Gyr are required to form an IMBH with mass $\sim{}500$ M$_\odot$ \cite{miller2002}.

Monte Carlo simulations by Giersz et al. (2015, \cite{giersz2015}) show that both the runaway collision scenario and the repeated merger scenario can be at work in star clusters: runaway collision IMBHs form in the first few Myr of the life of a star cluster and grow in mass very efficiently, while repeated-merger IMBHs start forming much later ($\gtrsim{}5$ Gyr) and their growth is less efficient. 

\subsection{Alternative models for massive BHs and IMBH formation in galactic nuclei}

Several additional models predict the formation of IMBHs in galactic nuclei. For example, Miller \& Davies  (2012, \cite{millerdavies2012}) propose that IMBHs can efficiently grow in galactic nuclei from runaway tidal capture of stars, provided that the velocity dispersion in the nuclear star cluster is $\gtrsim{}40$ km s$^{-1}$. Below this critical velocity, binary stars can support the system against core collapse, quenching the growth of the central density and leading to the ejection of the most massive BHs. Above this velocity threshold, the stellar density can grow sufficiently fast to enhance tidal captures and BH--star collisions. Tidal captures are more efficient than BH--star collisions in building up IMBHs, because the mass growth rate of the former scales as $\dot{m}_{\rm IMBH}\propto{m_{\rm IMBH}}^{4/3}$ (where $m_{\rm IMBH}$ is the IMBH seed mass), whereas the mass growth of the latter scales as $\dot{m}_{\rm IMBH}\propto{m_{\rm IMBH}}$ \cite{stone2017}. 

Furthermore, McKernan et al. (2012, \cite{mckernan2012}, see also \cite{mckernan2014,mckernan2018,bartos2017,yang2019a,yang2019b}) suggest that IMBHs could efficiently grow in the accretion disc of a SMBH. Nuclear star cluster members trapped in the accretion disk are subject to two competing effects: orbital excitation due to dynamical heating by encounters with other stars and orbital damping due to gas drag. Gas damping is expected to be more effective than orbital excitation, quenching the relative velocity between nuclear cluster members and enhancing the collision rate. This favours the growth of IMBHs via both gas accretion and multiple stellar collisions. This mechanism might be considered as a gas-aided hierarchical merger scenario. 

\vspace{0.5cm}

As a final remark, it is worth mentioning that all the IMBH formation scenarios we have discussed here -- i)runaway collisions of massive stars, ii) hierarchical merger of BHs and iii) BH trapping in AGN discs-- can also work as possible formation mechanisms for GW190521-like events: they can lead to the formation of BHs with mass in the pair instability mass gap and with large spins. These BHs are born in dense stellar environments, where they can acquire companions dynamically and form BBHs with isotropically oriented spins \cite{abbottGW190521astro}. 

\subsection{Kozai-Lidov resonance}

Unlike the other dynamical processes discussed so far, Kozai-Lidov (KL) resonance \cite{kozai1962,lidov1962} can occur both in the field and in star clusters. KL resonance appears whenever we have a stable hierarchical triple system (i.e., a triple composed of an inner binary and an outer body orbiting the inner binary), in which the orbital plane of the outer body is inclined with respect to the orbital plane of the inner binary. Periodic perturbations induced by the outer body on the inner binary cause i) the eccentricity of the inner binary and ii) the inclination between the orbital plane of the inner binary and that of the outer body to oscillate.  The semi-major axis of the binary star is not affected, because KL resonance does not imply an energy exchange between inner and outer binary. 
KL oscillations may enhance BBH mergers, because the timescale for merger by GW emission strongly depends on the eccentricity $e$ of the binary $t_{\rm GW}\propto{}(1-e^2)^{7/2}$ (see equation~\ref{eq:peters}, \cite{peters1964}).

It might seem that hierarchical triples are rather exotic systems. This is not the case. In fact, $\sim{}10$\% of low-mass stars are in triple systems \cite{tokovinin2008,tokovinin2014,raghavan2010}. This fraction gradually increases for more massive stars \cite{duchene2013}, up to $\sim{}50$\% for B-type stars \cite{sana2014,moe2017,toonen2016}. In star clusters, stable hierarchical triple systems may form dynamically, via four-body or multiple-body encounters.

Kimpson et al. (2016, \cite{kimpson2016}) find that KL resonance may enhance the BBH merger rate by $\approx{}40$\% in young star clusters and open clusters. According to Fragione et al. (2019, \cite{fragione2019a}), the merger fraction of BBHs in galactic nuclei can be up to $\sim{}5-8$ times higher for triples than for binaries. On the other hand, Antonini et al. (2017, \cite{antonini2017}) find that KL resonance in field triples can account for $\lesssim{}3$ mergers Gpc$^{-3}$ yr$^{-1}$.  The main signature of the merger of a KL system is the non-zero eccentricity until very few seconds before the merger. Eccentricity might be significantly non-zero even when the system enters the LIGO--Virgo frequency range. 


KL resonances have an intriguing application in nuclear star clusters. If the stellar BH binary is gravitationally bound to the super-massive BH (SMBH) at the centre of the galaxy, then we have a peculiar triple system where the inner binary is composed of the stellar BH binary and the outer body is the SMBH \cite{antonini2012}. Also in this case, the merging BBH has good chances of retaining a non-zero eccentricity till it emits GWs in the LIGO--Virgo frequency range.

\subsection{Summary of dynamics and open issues}

In this Section, we have seen that dynamics is a crucial ingredient to understand BBH demography. Dynamical interactions (three- and few-body close encounters) can favour the coalescence of BBHs through dynamical hardening. New BBHs can form via dynamical exchanges and via direct three-body encounters or GW captures. Hierarchical mergers of BHs in dense star clusters and in AGN disks can trigger the formation of unusually massive BHs (like GW190521) and even IMBHs. Stellar mergers might also lead to the formation of unusually massive BHs (like GW190521) and IMBHs in dense star clusters. All these dynamical processes suggest a boost of the BBH merger rate in dynamically active environments.

Overall, dynamically formed BBHs are expected to be more massive than BBHs from isolated binary evolution, with higher initial eccentricity and with misaligned spins. Also, KL resonances favour the coalescence of BBHs with extremely high eccentricity, even close to the last stable orbit. On the other hand, three-body encounters might trigger the ejection of binary compact objects from their natal environment, inducing a significant displacement between the birth place of the binary and the location of its merger.

\begin{figure}
\center{
\includegraphics[width=12cm]{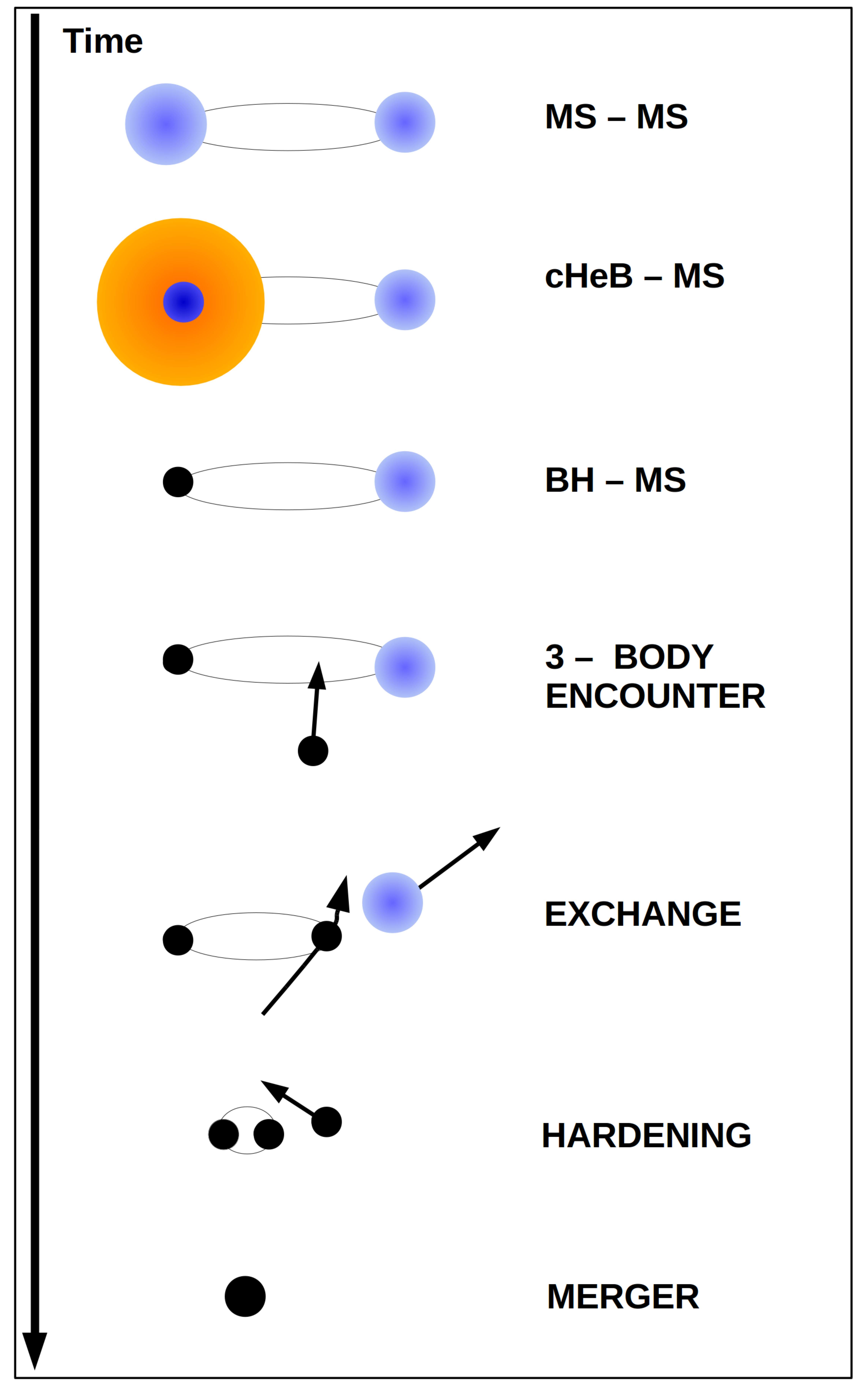}    
}
\caption{\label{fig:cartoonDYN} Schematic evolution of a merging BBH formed by dynamical exchange (see, e.g., \cite{downing2010,downing2011,ziosi2014,mapelli2016,rodriguez2015,rodriguez2016,askar2017}).
}
\end{figure}

Figure~\ref{fig:cartoonDYN} summarizes one of the possible evolutionary pathways of merging BBHs which originate from dynamics (the variety of this formation channel is too large to account for all dynamical channels mentioned above in a single cartoon). As in the isolated binary case, we start from a binary star. In the dynamical scenario, it is not important that this binary evolves through Roche lobe or CE (although this may happen). After the primary has turned into a BH, the binary undergoes a dynamical exchange: the secondary is replaced by a massive BH and a new BBH forms. The new binary system is not ejected from the star cluster and undergoes further three-body encounters. As an effect of these three-body encounters, the binary system hardens enough to enter the regime in which GW emission is efficient: the BBH merges by GW decay.

We expect dynamics to be important for BBHs also because massive stars (which are the progenitors of BHs) form preferentially in young star clusters \cite{portegieszwart2010}, which are dynamically active places. It is reasonable to expect that most BHs participate in the dynamics of their parent star cluster before being ejected by dynamical recoil or relativistic kicks. 

\section{BBHs in the cosmological context}

At design sensitivity, LIGO and Virgo will observe BBHs up to redshift $z\sim{}1$. The farthest event observed to date, GW190521, is at $z\sim{}0.8$, when the Universe was only $\sim{}6.8$ Gyr old. Next generation ground-based GW detectors, such as the Einstein Telescope \cite{punturo2010}, will observe BBH mergers up to redshift $z\gtrsim{}10$, when the Universe was $\lesssim{}500$ Myr old. Almost the entire Universe will be transparent to BBH mergers and it will be possible to investigate the cosmic evolution of stars and galaxies through the observations of BBH mergers. In other words, we will do cosmology through BBH mergers. 


Accounting for the cosmological context of BBH mergers  might appear as a desperate challenge, because of the humongous dynamical range: the orbital separations of BBHs are of the order of tens of solar radii, while cosmic structures are several hundreds of Mpc. Several theoretical studies have addressed this challenge, adopting two different methodologies. 

\subsection{Data-driven semi-analytic models}

Some authors (e.g. \cite{dominik2013,dominik2015,belczynski2016,lamberts2016,dvorkin2016,eldridge2016,dvorkin2018,giacobbo2018b,boco2019,neijssel2019,eldridge2019,baibhav2019,vitale2019,tang2020,santoliquido2020a,santoliquido2020b}) combine the outputs of population synthesis codes with analytic prescriptions. The main ingredients are the cosmic star formation rate density and the evolution of metallicity with redshift \cite{madau2014,madau2017,chruslinska2019a,chruslinska2019b,chruslinska2020}. In some previous work (e.g. \cite{dominik2013,lamberts2016}) a Press-Schechter like formalism is adopted, to include the mass of the host galaxy in the general picture. Lamberts et al. (2016, \cite{lamberts2016}) even include a redshift-dependent description for the mass-metallicity relation (hereafter MZR), to account for the fact that the mass of a galaxy and its observed metallicity are deeply connected. The main advantage of this procedure is that the star formation rate and the metallicity evolution can be derived more straightforwardly from the data. The main drawback is that it is more difficult to trace the evolution of the host galaxy of the BBH, through its galaxy merger tree.


In the semi-analytic description, the merger rate density evolution can be described as \cite{santoliquido2020a}:
\begin{equation}
\label{eq:mrd}
   \mathcal{R}(z) = \int_{z_{\rm max}}^{z}\psi(z')\,{}\frac{{\rm d}t(z')}{{\rm d}z'}\,{}\left[\int_{Z_{\rm min}}^{Z_{\rm max}}\eta(Z) \mathcal{F}(z',z, Z)\,{}{\rm d}Z\right]\,{\rm d}z',
\end{equation}
where ${\rm d}t(z')/{\rm d}z'=(1+z')^{-1}\,{}H(z')^{-1}$, with $H(z')=H_0\,{}\left[(1+z')^3\,{}\Omega_{\rm M}+\Omega_\Lambda\right]^{1/2}$. Furthermore, $Z_{\rm min}$ and $Z_{\rm max}$ are the minimum and maximum metallicity, $\psi{}(z')$ is the cosmic star formation rate (SFR) at redshift $z'$, $\mathcal{F}(z',z,Z)$ is the rate  of compact binaries that form at redshift $z'$ from stars with metallicity $Z$ and merge at redshift $z$, and $\eta(Z)$ is the merger efficiency, namely the ratio between the total number $\mathcal{N}_{\rm TOT}(Z)$ of compact binaries (formed from a coeval population) that merge within an Hubble time ($t_{{\rm H}_0} \lesssim 14$ Gyr) and the total initial mass $M_\ast{}(Z)$ of the simulation with metallicity $Z$. The value of  $\mathcal{F}(z',z,Z)$ and that of  $\eta(Z)$ can be calculated either from catalogues of population-synthesis simulations or from phenomenological models.

\begin{figure}
\center{
\includegraphics[width=12cm]{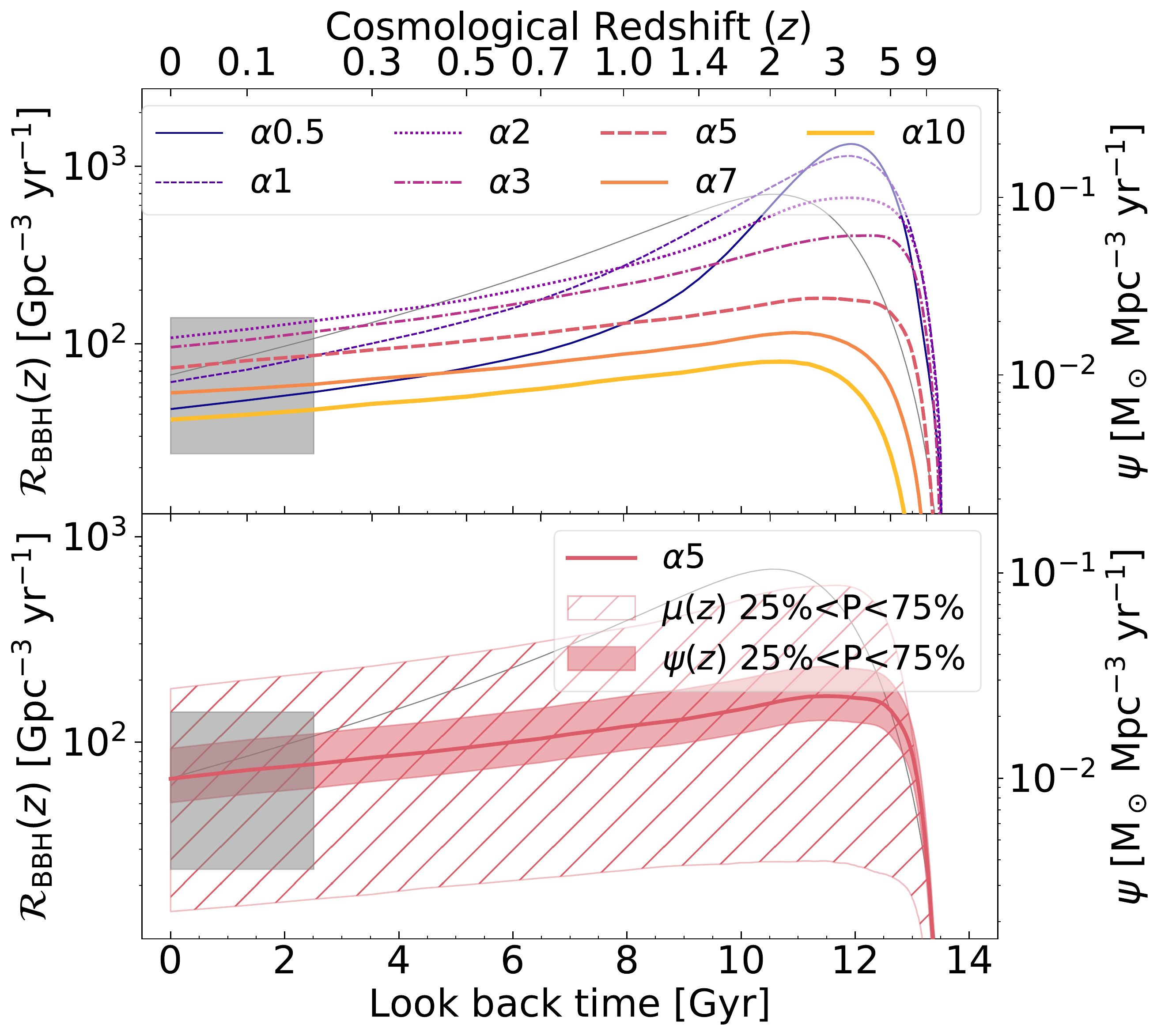}     
\caption{\label{fig:cosmicrate} Cosmic merger rate density of BBHs in the comoving frame ($\mathcal{R}_{\rm BBH}$) as a function of the look-back time ($t_{\rm lb}$, lower $x-$axis) and of the redshift ($z$, upper $x-$axis) from some of the models presented in \cite{santoliquido2020b}, which are based on equation~\ref{eq:mrd}. We use catalogues of population synthesis simulations run with {\sc mobse} \cite{mapelli2017,giacobbo2018a}. In both panels: thin gray line and right $y-$ axis: cosmic star formation rate density as a function of redshift from \cite{madau2017}. Gray shaded box: 90\% credible interval for the local merger rate density of BBHs, as inferred from the LVC data, considering the union of the rates obtained with model A, B and C in \cite{abbottO2popandrate}. The width of the grey shaded area on the $x-$axis  corresponds to the instrumental horizon obtained by assuming BBHs with total mass of $20$ M$_\odot$ and O2 sensitivity \protect\cite{abbott2018observingscenario}.  Upper panel: BBH merger rate density for different assumptions on the $\alpha$ parameter (i.e. the efficiency of CE ejection). Models for $\alpha{}=0.5,$ 1, 2, 3, 5, 7 and 10 are shown. Lower panel: the solid red line shows the median value of the BBH merger rate density for $\alpha{}=5$, while the shaded areas show the 50\% credible areas estimated from the uncertainties on metallicity evolution (hatched region) and star formation rate density evolution (shaded region). The cosmic star formation rate density evolution is modelled as in \cite{madau2017} and is obtained by fitting data. The metallicity evolution uses the fitting formula by \cite{decia2018}, adapted as decribed in \cite{santoliquido2020a}.  Courtesy of Filippo Santoliquido.}}
\end{figure}

This formalism yields an evolution of the merger rate density with redshift as shown in Fig.~\ref{fig:cosmicrate}. From the upper panel, we see that the cosmic merger rate density evolution is sensitive to the choice of the CE parameter $\alpha$. From the lower panel, it is apparent that the BBH merger rate density is tremendously affected by the metallicity evolution and by the observational uncertainties on metallicity evolution. Most models agree on this result (e.g., \cite{boco2019,neijssel2019,eldridge2019,baibhav2019,tang2020,santoliquido2020a,santoliquido2020b}). The reason for this trend is twofold. On the one hand, according to current models of BBH formation, the merger efficiency $\eta{}$ is 2$-$4 orders of magnitude higher at low metallicity ($Z<0.0002$) than at high metallicity. Hence, the merger rate is extremely sensitive to the underlying metallicity evolution. On the other hand, the uncertainties on metallicity evolution from observational data are large, as discussed by, e.g., \cite{madau2014,maiolino2019,chruslinska2019a,chruslinska2019b,santoliquido2020a,chruslinska2020}. 

The trend of the cosmic merger rate in Fig.~\ref{fig:cosmicrate} is similar to the trend of the cosmic star formation rate density curve, modulated by both metallicity evolution and time delay\footnote{We define as time delay the time elapsed between the formation of the progenitor binary and the merger of the two BHs.}. The peak of the BBH merger rate density curve ($z\sim{}3-4$, depending on $\alpha$)  is at a higher redshift than the peak of the cosmic star formation rate density ($z=2$). This happens because the lower the metallicity is, the  higher is the efficiency of BBH mergers, and metal-poor stars are more common at high redshift than at low redshift. With this approach, it will also be possible to put constraints, in a statistical sense, on the fraction of BBH mergers from the dynamical channel with respect to BBH mergers from the isolated evolutionary channel (e.g., \cite{rodriguez2018,bouffanais2019,santoliquido2020a,wong2020,zevin2020b}).

\subsection{Cosmological simulations}

The alternative approach to reconstruct the BBH merger history feeds the outputs of population-synthesis simulations into cosmological simulations \cite{oshaughnessy2017,schneider2017,mapelli2017,mapelli2018,mapelli2018b,mapelli2019,cao2018,marassi2019,graziani2020,artale2019,artale2020a,artale2020b}, through a Monte Carlo approach. This has the clear advantage that the properties of the host galaxies can be easily reconstructed across cosmic time. However, the ideal thing would be to have a high-resolution cosmological simulation (sufficient to resolve also small dwarf galaxies) with a box as large as the instrumental horizon of the GW detectors. This is obviously impossible. High-resolution simulations have usually a box of a few comoving Mpc$^3$, while simulations with a larger box cannot resolve dwarf galaxies. Moreover, this procedure requires to use the cosmic star formation rate density and the redshift-dependent MZR which are intrinsic to the cosmological simulations. While most state-of-the-art cosmological simulations reproduce the cosmic star formation rate density reasonably well, the MZR is an elusive feature, creating more than a trouble even in the most advanced cosmological simulations.


For example, \cite{mapelli2018b} investigate the main properties of the host galaxies of merging BHs, by combining their population-synthesis simulations \cite{mapelli2017,mapelli2018} with the {\sc illustris} cosmological box \cite{vogelsberger2014a,vogelsberger2014b,nelson2015}. The size of the {\sc illustris} (length $=106.5$ comoving Mpc) is sufficient to satisfy the cosmological principle, but galaxies with stellar mass $\lesssim{}10^8$ M$_\odot$ are heavily under-resolved. Their results show that BHs merging in the local Universe ($z<0.024$) have formed in galaxies with relatively small stellar mass ($<10^{10}$ M$_\odot$) and relatively low metallicity ($Z\leq{}0.1$ Z$_\odot$). These BHs reach coalescence either in the galaxy where they formed or in larger galaxies (with stellar mass up to $\sim{}10^{12}$ M$_\odot$). In fact, most BHs reaching coalescence in the local Universe appear to have formed in the early Universe ($\gtrsim{}9$ Gyr ago), when metal-poor galaxies were more common. A significant fraction of these high-redshift metal-poor galaxies merged within larger galaxies before the BBHs reached coalescence by GWs. Moreover, these models show that the mass spectrum and the other main properties of BBHs do not evolve significantly with redshift \cite{mapelli2019}.

Schneider et al. (2017, \cite{schneider2017}, see also \cite{marassi2019,graziani2020}) adopt a complementary approach to study the importance of dwarf galaxies for GW detections. They use the {\sc gamesh} pipeline to produce a high-resolution simulation of the Local Group (length $= 4$ Mpc comoving). This means that the considered portion of the Universe is strongly biased, but the resolution is sufficient to investigate BH binaries in small ($\gtrsim{}10^6$ M$_\odot$) dwarf galaxies. One of their main conclusions is that GW150914-like events originate mostly from small metal-poor galaxies.

Similarly, Cao et al. (2018, \cite{cao2018}) investigate the host galaxies of BBHs by applying a semi-analytic model to the Millennium-II N-body simulation \cite{boylan2009}. The Millennium-II N-body simulation is a large-box (length = 137 comoving Mpc) dark-matter only simulation. The physics of baryons is implemented  through a semi-analytic model. Using a dark-matter only simulation coupled with a semi-analytic approach offers the possibility of improving the resolution significantly, but baryons are added only in post-processing. With this approach, Cao et al. (2018, \cite{cao2018}) find that BBHs merging at redshift $z\lesssim{}0.3$ are located mostly in massive galaxies (stellar mass $\gtrsim{}2\times{}10^{10}$ M$_\odot$). 

Finally, \cite{artale2019,artale2020a,artale2020b} combine population-synthesis simulations with the {\sc eagle} cosmological simulation \cite{schaye2015}. They show that the merger rate per galaxy strongly correlates with the stellar mass of a galaxy for both BBHs, BHNSs and BNSs. They also find a dependence of the merger rate per galaxy on the star formation rate and a weaker dependence on the metallicity, but the correlation with the stellar mass is definitely the strongest one. This result has been used to optimize electromagnetic follow-up strategies, weighting the galaxies in the LVC uncertainty box by their stellar mass \cite{ducoin2020,ackley2020}. These studies show that the combination of population-synthesis tools with cosmological simulations is an effective approach to understand the cosmic evolution of the merger rate and the properties of the host galaxies of BBH mergers. 



\section{Summary and outlook}

We reviewed our current understanding of the formation channels of BBHs. The last five years have witnessed considerable progress in this field, thanks to the ground-breaking results of the LVC and to a renewed interest in BH astrophysics, mostly triggered by GW data.

The final stages of massive star evolution deserve particular attention to understand the mass and spins of BHs. GW data have recently challenged the concept of pair-instability mass gap, i.e. the existence of a mass range (between $\sim{}65$ M$_\odot$ and $\sim{}120$ M$_\odot$) in which we do not expect to find BHs as an effect of (pulsational) pair instability.

The process of mass transfer between two massive stars in a binary system is another key aspect shaping the demography of BBHs. The efficiency of mass accretion via Roche lobe overflow and the stability of mass transfer are possibly the main unknowns affecting the mass and the delay time of a BBH. In addition, the dynamical formation channel with its richness and multifaceted processes (hardening, exchanges, ejections, runaway collisions, hierarchical mergers, etc) adds to the complexity of the general picture.

Current astrophysical models of BBH formation face a number of essential questions:
\begin{itemize}
\item{}impact of core overshooting, rotation and possibly magnetic fields on massive star evolution and BH formation;
\item{}angular momentum transfer in massive stars and its link to BH spins;
\item{}explodability of massive stars;
\item{}stability and efficiency of mass transfer;
\item{}evolution of common-envelope systems;
\item{}physics of stellar collisions and mergers;
\item{}impact of the environment (e.g., metallicity, star clusters versus field) on the formation of BBHs across cosmic time.
\end{itemize}

Overall, we have a better understanding of the main dynamical processes with respect to both massive star evolution and SN physics. But our understanding of BBH dynamics is mainly hampered by two problems: i) we cannot properly model the dynamics of BBHs if we do not know BH masses and spins from stellar evolution; ii) dynamical simulations are still too computationally expensive to allow us to investigate the relevant parameter space.

Next-generation ground-based detectors will observe BBH mergers up to redshift $z\gtrsim{}10$, beyond the epoch of cosmic reionization \cite{kalogera2019}. This will open a completely new scenario for the study of BBHs across cosmic time and for the characterization of their evolutionary pathways.

\begin{acknowledgement}
We thank the DEMOBLACK team for useful discussions and for providing us with some essential material for this review.  MM  acknowledges financial support from the European Research Council for the ERC Consolidator grant DEMOBLACK, under contract no. 770017.
 \end{acknowledgement}

\bibliography{bibliography}
\end{document}